\documentclass{elsarticle}
\usepackage[english]{babel}
\usepackage{tabularx}
\usepackage{amsmath, amsfonts, amssymb, bm} 
\usepackage{graphicx}
\usepackage{siunitx}
\usepackage[letterpaper,top=2cm,bottom=2cm,left=3cm,right=3cm,marginparwidth=1.75cm]{geometry}
\usepackage{esvect}
\usepackage{makecell} 
\usepackage{float}
\usepackage{subcaption} 
\usepackage[table]{xcolor}
\usepackage{booktabs}
\usepackage{xcolor}
\usepackage{algpseudocode}
\usepackage{algorithm}
\usepackage[colorlinks=true, allcolors=blue]{hyperref}
\usepackage{tikz}
\usepackage{array}
\usepackage{hhline}
\usepackage{xcolor}   
\definecolor{headercolor}{RGB}{63,81,181}
\definecolor{rowcolor1}{RGB}{224,235,255}
\definecolor{rowcolor2}{RGB}{240,240,240}
\usepackage{comment}

\begin{document}
\begin{frontmatter}
\title{
Probabilistic
Predictions of Process-Induced Deformation in Carbon/Epoxy Composites Using a Deep Operator Network}
\author[1]{Elham Kiyani}
\author[2,3]{Amit Makarand Deshpande}
\author[7]{Madhura Limaye}
\author[1]{Zhiwei Gao}
\author[1]{Zongren Zou}
\author[2]{Sai Aditya Pradeep}
\author[2,3,4,5,6]{Srikanth Pilla}
\author[7]{Gang Li}
\author[7]{Zhen Li}
\author[1]{George Em Karniadakis}

\address[1]{Department of Applied Mathematics, Brown University, USA}
\address[2]{Center for Composite Materials, University of Delaware, Newark, Delaware 19716, USA}
\address[3]{Department of Mechanical Engineering, University of Delaware, Newark, Delaware 19716, USA}
\address[4]{Department of Material Science and Engineering, University of Delaware, Newark, Delaware 19716, USA}
\address[5]{Department of Chemical and Biomolecular Engineering, University of Delaware, Newark, Delaware 19716, USA}
\address[6]{Department of Computer and Information Sciences, University of Delaware, Newark, Delaware 19716, USA}
\address[7]{Department of Mechanical Engineering,
Clemson University, Clemson, SC 29634, USA}

\begin{abstract}
Fiber reinforcement and polymer matrix respond differently to manufacturing conditions due to mismatch in coefficient of thermal expansion and matrix shrinkage during curing of thermosets. These heterogeneities generate residual stresses over multiple length scales, whose partial release leads to process-induced deformation (PID), requiring accurate prediction and mitigation via optimized non-isothermal cure cycles. This study considers a unidirectional AS4 carbon fiber/amine bi-functional epoxy prepreg and models PID using a two-mechanism framework that accounts for thermal expansion/shrinkage and cure shrinkage. The model is validated against manufacturing trials to identify initial and boundary conditions, then used to generate PID responses for a diverse set of non-isothermal cure cycles (time-temperature profiles). Building on this physics-based foundation, we develop a data-driven surrogate based on Deep Operator Networks (DeepONets). A DeepONet is trained on a dataset combining high-fidelity simulations with targeted experimental measurements of PID. We extend this to a Feature-wise Linear Modulation (FiLM) DeepONet, where branch-network features are modulated by external parameters, including the initial degree of cure, enabling prediction of time histories of degree of cure, viscosity, and deformation. Because experimental data are available only at limited time instances (for example, final deformation), we use transfer learning: simulation-trained trunk and branch networks are fixed and only the final layer is updated using measured final deformation. Finally, we augment the framework with Ensemble Kalman Inversion (EKI) to quantify uncertainty under experimental conditions and to support optimization of cure schedules for reduced PID in composites.

\end{abstract}


\end{frontmatter}
\section{Introduction}
Thermoset fiber reinforced composites are widely adopted for lightweight structural applications in aerospace~\cite{1_Sudhin2020}, wind energy~\cite{2_murray2021}, automotive\cite{3_Deshpande2022,4_deshpande2023,5_pradeep2024}, prosthetics, wearable electronics and sports equipment. A common route for their manufacturing is the lay-up of pre-impregnated sheets of unidirectional fibers and subsequently curing them under elevated pressure and temperature. During this process, residual stresses are induced due to the mismatch between the thermal expansion coefficients of the fiber and the matrix, as well as the shrinkage of the thermoset matrix as it cures and cross-links, thus affecting the mechanical properties of the manufactured component~\cite{6_White1993,7_Li2014,8_Zhao2006,9_Agius2016,10_Hahn1976}. Release of some of these residual stresses due to material compliance and demolding results in process-induced deformation (PID) that can render the manufactured part out of design tolerances and defective. Additional stresses may also be induced if the deformed and out-of-tolerance part is assembled and fastened in place, potentially leading to premature or catastrophic failure. Accounting for PID and designing the cure cycle for PID reduction is therefore a critical aspect of the design and manufacturing of composite parts. Residual stresses have been shown to reduce by as much as 25--30\% through optimization of the cure cycle~\cite{25_White1993}.
Several studies have tried to develop predictive capabilities using numerical methods. Early analytical and numerical studies also examined the cured shape and process-induced deformation of composite shell structures, including cross-ply composite thin shells~\cite{ren2003cured}.
Constitutive laws have been proposed and demonstrated as part of finite element schemes to model cure dependent viscosity development and evolution of modulus and corresponding residual stresses~\cite {11_Li2018,12_Ding2016,13_Liu2021,14_Zobeiry2010}. White et al.~\cite{6_White1993} investigated, both experimentally and analytically (on LamCure), the effect of changing the manufacturer recommended cure cycle on the resulting curvature for asymmetric cross-ply laminates. Li et al.~\cite{17_Li2021} utilized a FE model for the multi-physics simulation to compute residual stresses, which was then modulated by an improved genetic algorithm to inversely determine the optimal process parameters to minimize residual stresses. Wang et al.~\cite{15_Wang2022} performed a fully coupled thermomechanical analysis to model the influence of cure cycles on residual stresses and degree of cure (DOC). They inferred that the temperature and duration of the second dwell had the greatest effect on residual stresses. Hui et al.~\cite{16_Hui2022} developed a collaborative multi-objective optimization strategy combining finite element-based cure process analysis with NGSA-II to demonstrate that the temperature gradient, residual stress, and process time can be reduced simultaneously. 

Kawagoe et al.~\cite{kawagoe2022multiscale} proposed a multiscale modeling framework, spanning quantum chemistry, molecular dynamics, and micro/macro finite-element analysis, to predict cure-induced deformation in CFRP laminates and demonstrated good agreement with fabrication experiments. Kinugawa et al.~\cite{kinugawa2025multiscale} studied process-induced residual deformation in CFRP laminates with reaction-induced phase-separated matrix resins and showed, through experiments and multiscale modeling, that thermoplastic resin addition reduces cure- and thermal-shrinkage-driven deformation. Saito et al. \cite{Saito2021,saito2021decoupling} developed a two-scale decoupled analysis framework for fiber reinforced plastics for unit cells composed of polymer resin matrix and carbon fibers. The proposed model based on the orthotropic version of the generalized Maxwell model, predicts the non-mechanical deformation associated with the interplay of cure shrinkage and thermal expansion/contraction. Thus, the model is able to simulate micro- and macroscopic thermomechanical responses of fiber reinforced plastics during and after cure. Humfeld et al.~\cite{humfeld2021machine} developed a machine-learning framework for real-time inverse modeling and multi-objective process optimization of composites for active manufacturing control, highlighting the growing role of data-driven methods in manufacturing-process monitoring and optimization.

Along with advances in physics-based modeling, rapid progress in machine learning (ML) has transformed manufacturing research by enabling efficient data-driven modeling, prediction, and optimization~\cite{behbahani2022machine,ademujimi2017review,ravanbakhsh2023combining,kiyani2023designing}. 
Deep learning, in particular, has become a central tool across science and engineering. Among these developments, operator learning has emerged as a powerful paradigm for approximating mappings between input and output functions, a setting that naturally arises in many scientific and engineering problems.
For PID prediction in composite structures, a range of ML architectures has been explored. Lin et al.~\cite{lin2021use} used neural networks, support vector regression, and $k$-nearest neighbors models to rapidly estimate differences in degree of cure from process parameters, while Fan et al.~\cite{fan2023deep} employed convolutional neural networks to predict PID fields for laminates with varying stacking sequences. Theory-guided and simulation-informed approaches have also been proposed, including Gaussian-process-based process optimization~\cite{schoenholz2024accelerated}, random-forest surrogates for optimal cure design in co-cured composite structures~\cite{2022_Lavaggi}, and simulation frameworks that capture competing mechanisms of residual-stress evolution under non-isothermal cure cycles~\cite{18_Limaye2024}. More recently, Liu et al.~\cite{liu2024rapid} combined viscoelastic finite element analysis, feature selection, and artificial neural networks for rapid PID prediction, while Zhang et al.~\cite{zhang2025deep} introduced deep-learning models tailored to complex thermo--chemo--mechanical interactions. Sequence models have also shown promise for curing-process prediction and real-time control; for example, Tang et al.~\cite{wenyuan2025real} combined a FE-based dataset, an LSTM network, and Q-learning to optimize curing trajectories for reduced thermal nonuniformity, cure nonuniformity, and tool--part interaction effects. Humfeld et al. \cite{Humfeld2021} utilized two recurrent neural networks to address the issue of uncertainty and variation in oven temperature distribution during large-scale manufacturing of composites in autoclaves. Real time temperature control is thus achieved to ensure the composite components are subjected to the right processing conditions, thereby mitigating process induced defects such as under-cure, overheating and excessive residual stress evolution.
Despite these advances, most existing ML approaches are formulated either as scalar-response predictors or as stepwise sequence models. In contrast, operator learning is particularly attractive for the present problem because it learns the mapping from an entire cure-temperature history to full response histories. This enables efficient prediction of the complete temporal evolution of degree of cure, viscosity, and deformation, while also providing a natural foundation for transfer learning, uncertainty quantification, and downstream cure-profile optimization.

Operator learning provides a general ML framework for approximating mappings between function spaces rather than finite-dimensional vectors~\cite{lu2019deeponet,lu2024bridging,shih2024transformers,wan2025deepvivonet,toscano2025variational}. A prototypical architecture for this task is the Deep Operator Network (DeepONet)~\cite{lu2019deeponet}, which employs a branch--trunk structure, the branch network encodes the input function, and the trunk network parameterizes the dependence on spatial or temporal query points. Several alternative neural operator architectures have been proposed, including Fourier Neural Operators (FNO)~\cite{li2020fourier} and U-Net-based operator models~\cite{ronneberger2015u}, which mainly differ in how they represent and propagate functional information. Physics-informed variants of DeepONet incorporate PDE residuals and other physical constraints into the loss function, enabling physically consistent predictions in data-scarce regimes~\cite{wang2021learning,kiyani2025predicting}. The Two-step DeepONet~\cite{lee2024training,kiyani2025predicting} extends this idea by composing multiple operator stages, thus increasing expressivity for complex multiscale or highly nonlinear mappings while retaining an operator-centric viewpoint. 

Building on these developments, Feature-wise Linear Modulation (FiLM)~\cite{perez2018film}
is extended to operator learning. FiLM was originally introduced for visual reasoning and
has since been incorporated into architectures such as graph neural networks~\cite{brockschmidt2020gnn}.
In the present work, a FiLM-augmented DeepONet surrogate model is employed to reconstruct
the full histories of DoC, viscosity, and deformation, despite the fact that only the
final deformation at the end of the cure cycle is available experimentally. To address the
mismatch between the time discretization used in simulations and that available in experiments,
a transfer learning strategy~\cite{pan2020transfer} is adopted, in which the DeepONet is first
trained on high-fidelity simulation data and subsequently fine-tuned using experimental
measurements through a loss function that enforces agreement between the predicted and
measured final deformation.

Recent operator-learning studies have extended DeepONet to better handle time-dependent inputs through sequential branch architectures, such as the GRU/LSTM-based S-DeepONet~\cite{he2024sequential} and transformer-based sequential neural operator frameworks~\cite{liu2025sequential}. These approaches are particularly well suited to high-dimensional transient loading problems, where strong sequential inductive biases can improve the representation of path-dependent dynamics. In the present work, however, the branch input is not a spatially varying transient field, but a low-dimensional global cure-temperature history together with the initial degree of cure. The temporal dependence of the outputs is represented through the trunk network, which is queried at arbitrary time instances to recover the full histories of degree of cure, viscosity, and deformation. For this setting, we adopt an MLP-based branch with FiLM conditioning as a simpler and more robust architecture for the available simulation and experimental data, while also enabling transfer learning from simulations to sparse experimental observations and uncertainty-aware optimization.

While the FiLM-DeepONet framework provides accurate predictions for the evolution of DoC, viscosity, and deformation, performing reliable uncertainty quantification (UQ) for operator learning remains substantially more challenging~\cite{psaros2023uncertainty, pensoneault2025uncertainty, uqfno, wang2026114640, zou2025uncertainty}. The associated inference problem is typically a very high-dimensional Bayesian inverse problem, which renders gradient-based sampling methods prohibitively expensive. Classical Bayesian approaches such as Hamiltonian Monte Carlo (HMC)~\cite{neal2012bayesian} and the Laplace approximation~\cite{mackay1992practical} can, in principle, deliver high-quality uncertainty estimates, but they scale poorly, even when diagonal approximations of the Hessian are employed.
As a practical alternative, Lakshminarayanan et al.~\cite{lakshminarayanan2017simple} proposed deep ensembles trained with proper scoring rules, which can yield well-calibrated predictive distributions with only minor modifications to standard training pipelines, 
and have recently been adopted in the context of operator learning~\cite{psaros2023uncertainty, zou2024neuraluq}. Randomized prior methods, which are likewise ensemble-based, have also been extended to UQ for operator learning \cite{yang2022scalable}. However, these ensemble methods have been criticized for lacking a clear Bayesian interpretation~\cite{wilson2022evaluating,reich2013nonparametric,pearce2018uncertainty}. To help bridge this gap, Pearce et al.~\cite{pearce2020uncertainty} introduced an ensemble scheme based on approximate Bayesian inference, in which model parameters are regularized around samples from a prior-like distribution, leading to more interpretable uncertainty estimates.

Another promising direction is ensemble Kalman inversion (EKI), a gradient-free ensemble method that has been successfully combined with UQ in scientific machine learning~\cite{gao2025scalable,pensoneault2025uncertainty}. Owing to its favorable scaling and rapid convergence, EKI is particularly well suited to high-dimensional operator-learning problems. In this work, we therefore employ EKI to train DeepONet while simultaneously obtaining uncertainty estimates.
To further reduce the effective parameter dimension and improve computational efficiency,
conventional multilayer perceptrons (MLPs) are replaced with Chebyshev enhanced
Kolmogorov Arnold Networks (cKANs) within the DeepONet architecture. The cKANs employed
here follow a modified DeepOKAN design, in which Chebyshev polynomials are incorporated
into Kolmogorov Arnold Networks (KANs)~\cite{liu2024kan,faroughi2025scientific}
within a standard DeepONet framework. Inspired by Kolmogorov networks~\cite{sprecher2002space,faroughi2025scientific,mostajeran2025minpo,koppen2002training},
KANs are designed to adapt their activation patterns to the input data, thereby enabling more
efficient forward evaluations and, consequently, faster generation of ensemble samples.
The model developed as part of this study, thus couples operator learning with transfer learning by training a DeepONet surrogate on simulation data and refining it with experimental measurements to recover full deformation fields from limited observations, while EKI-based training simultaneously yields uncertainty estimates, providing experimental predictions that are both accurate and uncertainty-aware.

In light of the above, the present work makes three main contributions. First, it introduces a FiLM-conditioned DeepONet surrogate for predicting the full time histories of degree of cure, viscosity, and deformation from cure-temperature histories under varying initial cure states. Second, it combines operator learning with transfer learning so that a surrogate trained on simulation data can be refined using sparse experimental measurements. Third, it employs EKI as a practical and scalable framework for training and uncertainty quantification in a high-dimensional operator-learning setting.

This paper is organized as follows. 
In Section~\ref{sec:Experimental_Materials_and_Methods}, the experimental materials, methods, and manufacturing trials are presented.
Building on this foundation, Section~\ref{sec:Model-and-Simulations} details the simulation framework and datasets used to model the cure process. 
Section~\ref{sec:Operator-learning} then introduces the operator-learning perspective, presenting the DeepONet and FiLM-DeepONet architectures, together with the associated network design and training strategy. 
The resulting predictive capabilities for cure processing are assessed in Section~\ref{sec:DeepONet-Based_results}. 
To quantify uncertainty in these predictions, Section~\ref{sec:Uncertainty_Quantification} develops DeepONet ensembles for epistemic uncertainty and incorporates the EKI methodology within the DeepONet-based framework. 
Leveraging these ingredients, Section~\ref{sec:Optimization} focuses on the optimization of the cure temperature profile. 
Finally, Section~\ref{sec:Summary} summarizes the main findings and outlines directions for future research.

\section{Experimental Materials and Methods}\label{sec:Experimental_Materials_and_Methods}
This section describes the experimental setup, specimen preparation, and materials used in this study.
\subsection{Experimental Approach}
Asymmetric laminates inherently develop residual stresses during cure, resulting in measurable out-of-plane deformation at the end of the process~\cite{6_White1993}. To leverage this effect, an asymmetric \([0/90]\) cross-ply laminate configuration was selected. This layup provides a clear and reproducible manifestation of PID, enabling direct comparison between simulations and experiments.
The simulation framework used in this study captures the physical mechanisms governing residual stress evolution and the resulting PID. Using this framework, two optimized non-isothermal cure cycles were identified. Experiments were then conducted in which the baseline cure cycle, along with the two optimized cycles, were implemented on a laboratory-scale setup, and the resulting PID was measured. These experiments serve two purposes:  
(i) to validate the simulation model by confirming that the initial and boundary conditions correctly represent the physical process, and  
(ii) to provide experimental data that complement the larger simulation-generated dataset for DeepONet model development.

The laminate layup was carried out with particular care to ensure consistent ply orientation across all specimens, thereby reducing variability associated with fiber misalignment. Samples fabricated from the selected thermoset prepreg were cured under constant consolidation pressure using one baseline isothermal cycle and two optimized non-isothermal cycles aimed at mitigating PID. The temperature histories of the specimens were recorded in real time and subsequently used to predict the deformation history using a FiLM-enhanced DeepONet.

\subsection{Materials and Methods}

The material system used in this study is Hexply\textsuperscript{\textregistered} 3501-6 (formerly Hercules 3501-6), supplied by Hexcel Corporation (West Valley City, Utah, USA). It is a thermoset prepreg composite consisting of non-crimped unidirectional AS4 carbon fiber filaments produced from polyacrylonitrile (PAN) precursor fibers and impregnated with a 3501-6 amine-based epoxy matrix, with a fiber volume fraction of 42\%. Hexply\textsuperscript{\textregistered} 3501-6 is widely employed in structural aerospace applications with a service temperature limit of up to \SI{177}{\degreeCelsius}. The prepreg was stored at \SI{-18}{\degreeCelsius} to preserve its shelf life when not in use to prepare samples for experimental manufacturing trials.
The thermoset resin used in this prepreg is a B-stage epoxy resin, meaning it is partially cured during impregnation of the fibers by the manufacturer. Upon application of the manufacturer-recommended cure cycle, the cross-linking reaction resumes and proceeds toward full cure. Consequently, the initial degree of cure, $\mathrm{DoC}_{0}$, must be quantified to establish realistic initial conditions for the simulations. In this study, $\mathrm{DoC}_{0}$ was calculated using Equation~\eqref{eq:alpha}, ensuring consistency between the experimental characterization and the simulation inputs.

\begin{equation}
\alpha = \left( \frac{\Delta H_{\text{residual cure}}}{\Delta H_{\text{full cure}}} \right) \times 100,
\label{eq:alpha}
\end{equation}

where:

\begin{itemize}
    \item $\Delta H_{\text{residual cure}}$ = Enthalpy of reaction of the as-received B-stage resin,
    \item $\Delta H_{\text{full cure}}$ = Ultimate enthalpy of reaction of freshly mixed resin prior to prepreg manufacturing and use.
\end{itemize}

Two differential scanning calorimetry (DSC) methods are typically used to determine $\alpha_i$: (1) measuring the residual heat of reaction in the as-received material, or (2) assessing the shift in the glass transition temperature ($T_g$)~\cite{19_Sun2002}. In this study, the first method was adopted, wherein the residual heat of reaction was measured to determine $\alpha_i$, following the approach described by Sun et al.~\cite{19_Sun2002}.
Hargis et al.~\cite{20_Hargis2006} performed a detailed calorimetric analysis of AS4/3501-6 using DSC to characterize the cure kinetics. Since cure kinetics generally describe the behavior of the thermoset epoxy resin matrix, it is important to accurately determine the resin mass fraction in the prepreg when DSC samples are taken directly from fiber-reinforced prepregs~\cite{21_Kim2002}. A more common alternative is to obtain the thermoset epoxy system from the manufacturer as separate, unmixed components, mix them immediately prior to the DSC run, and perform DSC on the neat epoxy system~\cite{22_IlLee1982,23_Chern2002}. 
However, this latter approach yields the total heat of reaction for freshly mixed resin and may not accurately represent the state of the epoxy in the prepreg, where some prereaction has already taken place.  Consequently, several studies have instead performed DSC measurements directly on AS4/3501-6 prepreg and calculated the heat of reaction while compensating for the fiber mass fraction present in the DSC sample~\cite{21_Kim2002,23_Chern2002}. The values reported in the literature are summarized in Table~\ref{tab:resize_exp}.

\begin{table}[!htb]
\centering
\renewcommand{\arraystretch}{1.3} 
\resizebox{\textwidth}{!}{
\begin{tabular}{|c|c|c|c|c|c|}
\hline
\textbf{Reference} & \textbf{Year} & \textbf{Material Composition} & \textbf{Measurement Conditions} & \textbf{Enthalpy of Reaction $H_t$ (J/g)} & \textbf{Details on Pre-reaction} \\
\hline

Lee et al.~\cite{22_IlLee1982} & 1982 & 3501-6 Resin & @ 20 °C/min to 326.85 °C & $473.6 \pm 5.4$ & NA \\

Hou et al.~\cite{24_Hou1988} & 1988 & 3501-6 Resin & @ 20 °C/min & $502 \pm 21$ & NA \\

White et al.~\cite{25_White1993} & 1993 & \makecell{AS4 CF/3501-6 Prepreg \\ (42\% resin mass)} & @ 20 °C/min to 350 °C & \makecell{182.87 (prepreg) \\ 435 (comp. for resin)} & NA \\

Kim et al.~\cite{21_Kim2002} & 2002 & \makecell{AS4 CF/3501-6 Prepreg \\ (37.4\% resin mass)} & @ 20 °C/min to 350 °C & 433.7 (comp. for resin) & NA \\

Kim et al.~\cite{21_Kim2002} & 2002 & \makecell{AS4 CF/3501-6 Prepreg \\ (37.4\% resin mass)} & @ 2 °C/min to 350 °C & 456.7 (comp. for resin) & NA \\

Chern et al.~\cite{23_Chern2002} & 2002 & 3501-6 Resin & @ 20 °C/min & $508 \pm 19$ & \makecell{2\% cure \\ (manufacturer’s note)} \\

Hargis et al.~\cite{20_Hargis2006} & 2006 & 3501-6 Resin & @ 20 °C/min & $382.5 \pm 20$ & \makecell{Significant \\ pre-reaction reported} \\
\hline
\end{tabular}
}
\caption{Comparison of the enthalpy of reaction ($H_t$) for the 3501-6 resin system and AS4 CF/3501-6 epoxy prepreg under dynamic DSC conditions. The measurement method (Dynamic) is common to all studies.
}
\label{tab:resize_exp}
\end{table}

Based on the values reported in Table~\ref{tab:resize_exp} and the procedure in Equation~\eqref{eq:alpha}, $\mathrm{DoC}_{0}$ was estimated to lie within the range of 17.68\%--31.21\% when the measurement standard deviations were taken into account. Using the reported mean values, $\mathrm{DoC}_{0}$ was calculated to be 24.7\%. Accordingly, an initial condition of $\mathrm{DoC}_{0}=0.3$ (30\%) was assumed for the simulations. This approximation also accounts for additional curing that may have occurred during the several months between prepreg manufacture and its use in the experiments.

\begin{figure}[!htb]
\centering
\includegraphics[width=0.75\textwidth]{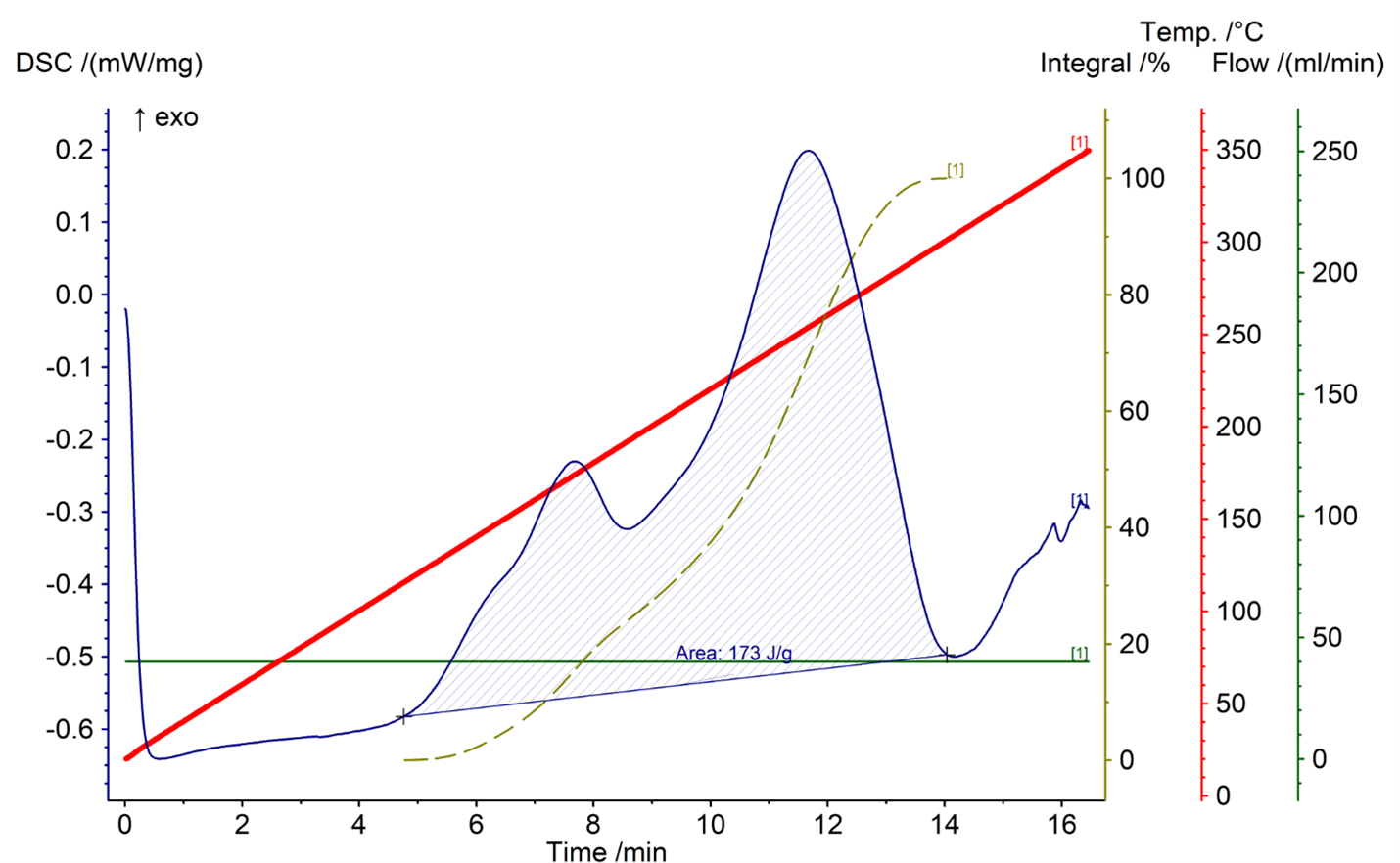}
\caption{Dynamic DSC run at heating rate of 20~°C/min indicating heat flow versus time for the AS4 CF/3501-6 epoxy prepreg, with the area under the peaks indicating the heat of reaction resulting from the curing of the resin due to heat input during the DSC run}
\label{fig:DSC_plot}
\end{figure}

The material used in the compression molding trials was characterized under comparable DSC conditions using a Netzsch Polyma 214 DSC (Netzsch Instruments Inc., Burlington, MA, USA), with nitrogen purge gas and an empty aluminum pan as reference. The results from the first dynamic heating run at 20~°C/min are presented in Figure~\ref{fig:DSC_plot} and show good agreement with the prepreg values reported by White et al.~\cite{25_White1993} (see Table~\ref{tab:resize_exp}).

\subsection{Manufacturing Trials}

\begin{figure}[!htb]
\centering         
\includegraphics[width=0.5\textwidth]{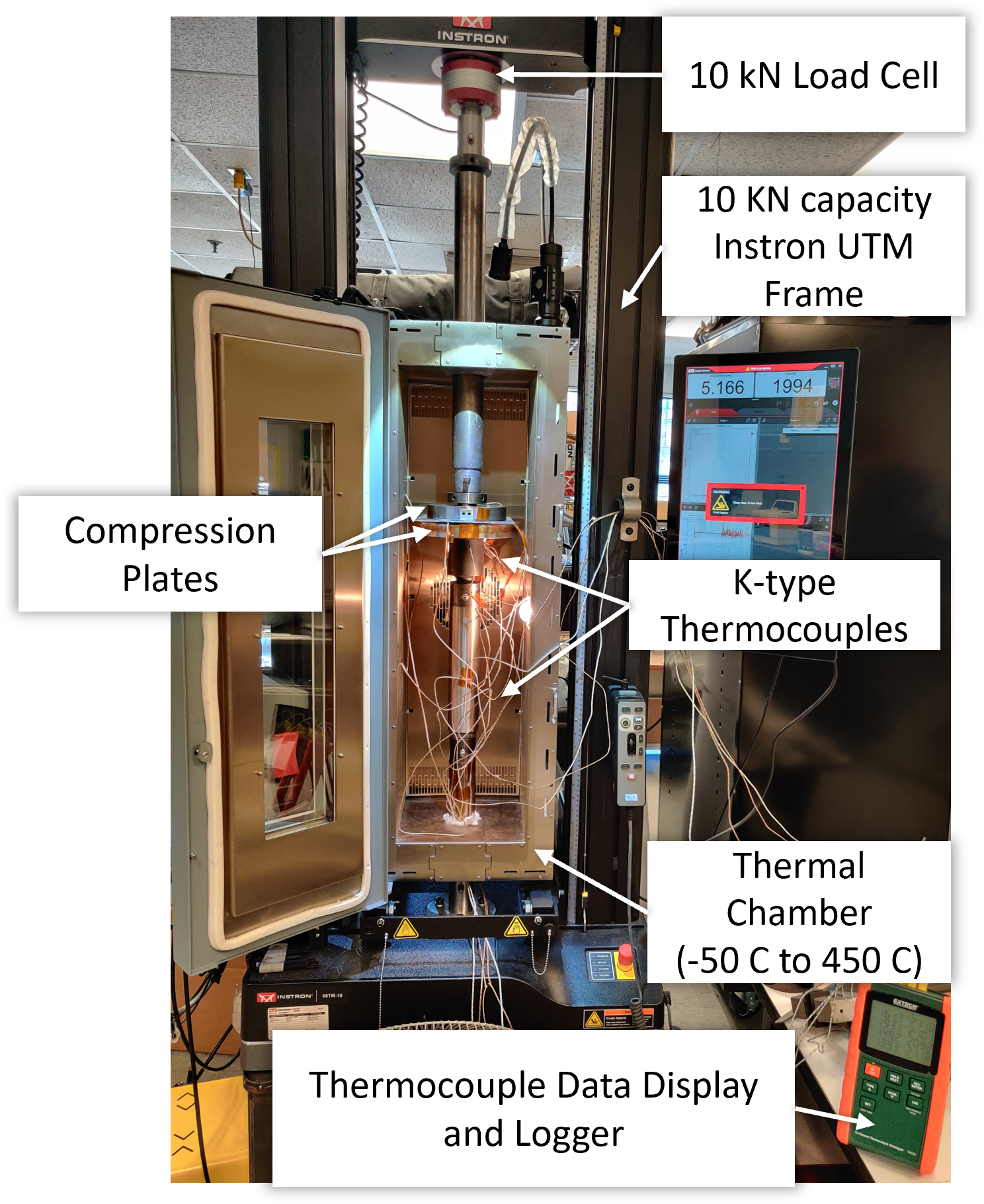}  
\caption{Experimental setup to consolidate and heat unbalanced [0/90] lay-up of prepregs to replicate the cure cycle to be followed during manufacturing, with a universal testing machine for precise displacement control of the compression plates, a 10 kN load cell measure and control the consolidfation force applied, and K-type thermocouples in contact with the material to measure and log temperature change.}  
\label{fig:Instron}  
\end{figure}

The experimental validation runs were performed on a lab-scale setup for compression molding. The setup utilized a 10 kN load capacity Instron 68TM-10 (Instron,Norwood, MA, USA) servo-electric universal testing machine (UTM) with a 10 kN load cell and a temperature chamber 3119-615 (Buckinghamshire, England) with a temperature range of -50 °C to 450 °C, capable of executing programmed heating and cooling cycles. The setup is as illustrated in Figure~\ref{fig:Instron}. 
It should be noted that all manufacturing trials were performed under the same consolidation force of 2 kN, corresponding to a consolidation of pressure of approximately 16-17 PSI, which is close to vacuum bagging pressure. 

Excessively high consolidation pressure influences achieved fiber volume fraction, matrix squeeze-out and thickness of the manufactured composite part that in turn affect process induced deformation. Thus, for the scope of this study, the process parameter of consolidation pressure has been kept constant. Furthermore, the value selected is both representative of a real-world composite manufacturing process utilizing prepregs, and low enough that the aforementioned aspects associated with excessively high pressure are mitigated.

\begin{figure}[!ht]
\centering         
\includegraphics[width=1\textwidth]{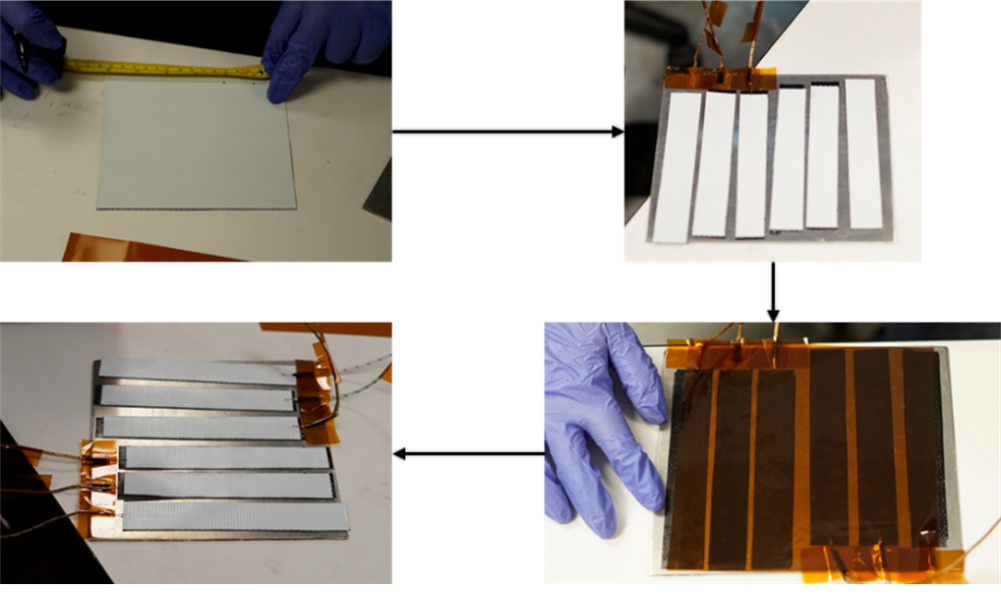}  
\caption{Specimen preparation methodology indicating layup of a 150 mm (6") x 150 mm (6") unbalance laminate followed by cutting into 6 specimens of 25.4 mm (1") width, arrangement of the specimens on the compression plates in contact with the thermocouples to enable in-situ temperature measurement and logging }  
\label{fig:preparation_exp}  
\end{figure}
The unbalanced ply layup of 152 mm (6”) x 152 mm (6”) unidirectional prepreg squares, were cut into 152 mm (6”) x 25.4 mm (1”) rectangular specimens, and placed onto a 3.17 mm (1/8th inch) thick aluminum caul plate. K-type thermocouples, with a measurement range of -50 °C to 400 °C, were placed in contact with the samples to accurately measure the material temperature. Additional thermocouples were also placed in contact with the top and bottom compression plates and the thermal chamber to ensure the thermal cycle was executed as programmed. Initial trials helped determine the correct heating rates, cooling rates and temperature setpoints to be programmed, so that the material was subjected to the right heating and cooling rates. Thus, the difference associated with the dynamic temperature lag between the chamber temperature and material temperature was accounted for. The specimen preparation is illustrated in Figure~\ref{fig:preparation_exp}. Henkel Loctite Frekote 55-NC, a semi-permanent, non-contaminating mold release agent was applied onto the compression plates to enable easy release of the cure laminates after curing.

\subsection{Experimental Cure Cycles and Measurements}\label{subsec:Experimental_Results}

\begin{figure}[!htb]
    \centering
    \includegraphics[width=0.99\textwidth]{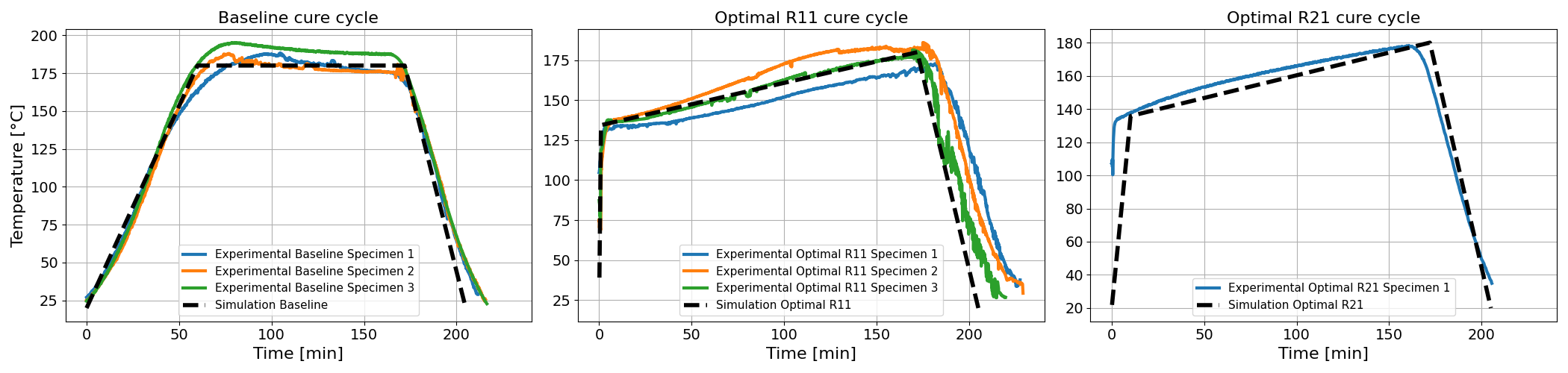}
    \caption{Temperature–time profiles for the three cure cycles: an isothermal baseline cycle and two non-isothermal cycles identified through the optimization scheme in~\cite{18_Limaye2024}. The curves show the measured specimen temperature, the thermal chamber environment, and the corresponding simulated cure profiles.}
    \label{fig:Experimentalcure_cycles_combined}
\end{figure}

\begin{figure}[!htb]
\centering         
\includegraphics[width=0.5\textwidth]{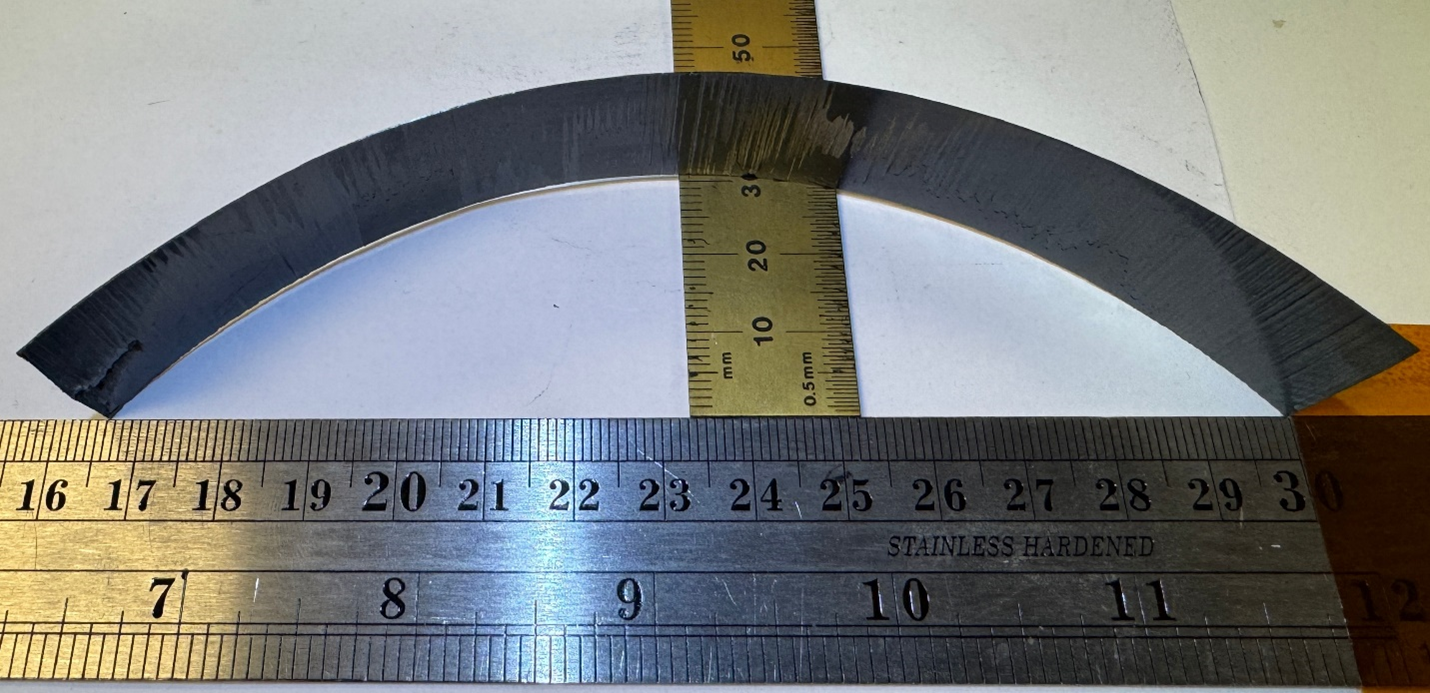}  
\caption{Measurement of process-induced deformation in unbalanced ply lay-up. The chord center point was identified and the perpendicular distance from that point to the curved unbalance laminate was measured}  
\label{fig:samplePID_measurement}  
\end{figure}

The specimens were subjected to three different cure cycles: one isothermal baseline cycle and two non-isothermal cycles identified through the optimization scheme developed  previously~\cite{18_Limaye2024}. The time–temperature histories were recorded and compared against the simulated cure profiles for the baseline, optimal R11, and optimal R21 cycles, as shown in Figure~\ref{fig:Experimentalcure_cycles_combined}.
After cooling to room temperature (\SI{20}{\degreeCelsius}), the caul plate and specimens were demolded. The cured laminates exhibited a curved profile analogous to a beam in bending, and the mid-span deflection was measured as illustrated in Figure~\ref{fig:samplePID_measurement}. 
and the resulting PID values were measured using this setup. These measurements were used to validate the simulation predictions and to assess the robustness of the optimization scheme aimed at minimizing PID. 
Multiple repeats were performed for the baseline and R11 cure cycle, with three to four specimens manufactured for each repeat. Manufacturing trial repeats were necessary for baseline and R11 cure cycle due to the deviations in specimen temperature measurements from the ideal simulated cure cycle temperatures. These deviations can be seen in Figure~\ref{fig:Experimentalcure_cycles_combined} when transitioning from a faster heating rate to an isothermal hold (in case of baseline) or when transitioning from a heating to cooling (in case of R11). This is attributed to the large volume within the thermal chamber which has a damping effect, preventing sharp transition points. This is a physical limitation of manufacturing with bulk materials, and is representative of the real-world manufacturing scenario for large unbalanced composite laminates. 

The manufacturing trial for R21 cure cycle, however, does not exhibit these fluctuations and hence did not necessitate multiple repeats, as is evident from Figure ~\ref{fig:Experimentalcure_cycles_combined}. A high heating rate (such as the one seen for R21 cycle) can be achieved with this setup by heating the chamber prior to inserting the prepreg specimens for consolidation pressure application and cure. Thus, additional repeats have been performed and reported for baseline and R11 cure cycle manufacturing trials to account for process variability and its implications on PID.
The experimental PID measurements and corresponding simulation results are also summarized in Table~\ref{tab:pid_validation}.  The variation between specimens for the same manufacturing trial, represented by the standard deviation for experimental measurements for PID can be attributed to variation the temperature distribution inside the thermal chamber and along the surface of the caul plate, based on the air flow characteristics within the thermal chamber. This temperature variability is a known phenomenon in manufacturing of laminated composites in bulk quantities, with active research efforts underway to mitigate its effects and develop novel AI/ ML assisted process control strategies around it \cite{Humfeld2021}. 

The simulated PID for the baseline cure cycle lies within the experimental standard deviation, while the simulations for the optimal cycles exhibit less than 10\% error relative to the measured values. This level of agreement provides confidence in the validated simulations and supports the use of additional optimization runs, performed with the objective of minimizing PID, as reliable data for subsequent AI model development.

\begin{table}[!htb]
    \centering
    \renewcommand{\arraystretch}{1.35} 
    \setlength{\tabcolsep}{2.2pt}     
    \small
    \begin{tabular}{|l|c|c|c|c|c|c|c|c|}
        \hline
        \makecell[l]{Thermal\\cycle} & Run & \makecell[c]{Specimen 1\\PID (mm)} & \makecell[c]{Specimen 2\\PID (mm)} & \makecell[c]{Specimen 3\\PID (mm)} & \makecell[c]{Specimen 4\\PID (mm)} & \makecell[c]{Mean PID\\(mm)} & \makecell[c]{Standard deviation\\(mm)} & \makecell[c]{Simulated PID\\(mm)} \\
        \hline
        Baseline    & 1 & 41.5 & 35.0 & 37.0 & 41.5 & 38.750 & 3.278 & 39.23 \\
        Baseline    & 2 & 39.0 & 36.5 & 34.5 & --   & 36.667 & 2.254 & 39.23 \\
        Baseline    & 3 & 36.0 & 41.0 & 36.0 & 36.5 & 37.375 & 2.428 & 39.23 \\
        \hline
        Optimal R11 & 1 & 36.5 & 37.0 & 37.5 & --   & 37.000 & 0.500 & 36.64 \\
        Optimal R11 & 2 & 33.5 & 35.0 & 35.0 & --   & 34.500 & 0.866 & 36.64 \\
        Optimal R11 & 3 & 31.0 & 33.5 & 36.0 & --   & 33.500 & 2.500 & 36.64 \\
        \hline
        Optimal R21 & 1 & 31.0 & 32.0 & 35.5 & --   & 32.833 & 2.362 & 36.84 \\
        \hline
    \end{tabular}
    \caption{Comparison of experimentally measured and simulated PID (mm) for the three cure cycles: an isothermal baseline cycle and two non-isothermal cycles identified through the optimization scheme in~\cite{18_Limaye2024}.}
    \label{tab:pid_validation}
\end{table}

\section{Model and Simulations}\label{sec:Model-and-Simulations}
A comprehensive study by Limaye et al.~\cite{18_Limaye2024} demonstrated that the cure cycle
governs phase transitions in polymer laminates, progressing from viscous to viscoelastic and
ultimately to elastic behavior. The study further highlighted the critical role of thermo-mechanical
interactions in modulus evolution and the development of residual stresses and deformations.
These coupled phenomena were represented using Polymerization--Gelation--Vitrification (PGV)
plots, which characterize material behavior as a function of the applied cure cycle.
To capture these complex interactions, a computational cure analysis framework was developed
using ABAQUS and COMPRO in a sequential two-step approach, comprising thermo-chemical
analysis followed by stress--deformation analysis. The model predicted temperature evolution,
DoC, and laminate deformation. These predictions were validated against existing
literature, showing strong agreement in both DoC profiles and deformation
measurements.

An optimization scheme was developed and implemented for the R11 and R21 cases using the
NSGA-II genetic algorithm, with the objective of minimizing PID.
This was accomplished by adjusting the slopes of the cure cycle’s two linear ramp segments, resulting in non-isothermal cure cycles that reduced overall deformation by balancing cure shrinkage with thermal expansion during processing. For these cases, $\mathrm{DoC}_{0}$ was prescribed, while the heating time was fixed at 1~minute for R11 and 10~minutes for R21. The optimization was performed subject to the constraints of achieving a high final cure state ($\mathrm{DoC} \ge 0.990$) and promoting interaction between thermal expansion and chemical shrinkage effects. 

The thermo-chemical and stress–deformation analyses in this work are based on the Cure-
Hardening Instantaneously Linear Elastic (CHILE) model, wherein the material is treated as an
elastic solid with stiffness that increases monotonically with the DOC. The glass
Tg is coupled to DOC through the DiBenedetto relation, while thermal
expansion and chemical shrinkage are incorporated via a DOC-dependent elastic constitutive law.
The evolution of DOC and viscosity is modeled following Woo et al.~\cite{lee1982heat}. Detailed formulations and
implementation of the CHILE model can be found in prior studies~\cite{limaye2025numerical,shah2018optimal,johnston1997integrated}. This approach does not
account for viscoelastic stress relaxation during curing and therefore tends to overpredict residual
stresses and PID. Additional assumptions include an approximate
treatment of viscosity, particularly post-gelation, and the exclusion of vitrification effects in the cure
kinetics, which may affect late-stage DOC evolution. Accordingly, the results should be interpreted
within these modeling assumptions. The scope of the analysis is to examine the influence of process parameters (e.g., temperature and cure time) on PID using a computationally efficient modeling
approach, and to generate training data for the DeepONet framework. The
proposed framework is cure model-agnostic and can be extended in future work to incorporate
higher-fidelity constitutive models, including viscoelasticity and multi-scale modeling approaches~\cite{chen2019improved,ryuzono2025mechanism}.

The mechanical model was formulated by applying displacement boundary conditions to eliminate rigid body motion. Specifically, the edge at (0,y,0) was fully constrained in all displacement directions, while the opposite edge at (a,y,0) was constrained in the y and z directions but allowed to slide freely along the x-direction, ensuring numerical stability without over-constraining the laminate. The explicit application of compaction pressure and detailed contact interactions between the laminate and tooling (caul and compression plates) were not included. This simplification is supported by prior work by Shah et al.~\cite{shah2018optimal}, which demonstrated that compaction pressure has a negligible effect on PID for similar laminate systems. The validity of this assumption is further supported by the good agreement between the model-predicted PID and experimental measurements obtained in the present study.

The reduction in PID under the optimized time–temperature conditions is governed by the thermochemical and thermomechanical response during cure, as established in prior work~\cite{limaye2024use}. The evolution of deformation is controlled by the interaction between thermal expansion and cure-induced chemical shrinkage, which are functions of the DoC evolution. The prescribed cure cycle influences the timing and magnitude of these mechanisms. For cases R11 and R21 considered in the present study, it was observed that increasing the lower bound of the initial heating time t from 1 min to 10 min leads to an increase in PID. This indicates that, for a fixed total cure time, longer initial heating durations promote higher deformation due to the reduced duration of interplay between thermal expansion and cure shrinkage. Consequently, from both a physical and manufacturing standpoint, minimizing the initial heating time is beneficial for reducing PID.

In the present work, the cure model is regenerated to evaluate non-isothermal cure cycles while accounting for material pre-cure through the prescribed value of $\mathrm{DoC}_{0}$. The end-of-cycle deformation predicted for the simulated cure cycles R11 and R21 is compared against experimental measurements (see Table~\ref{tab:pid_validation} and Figure~\ref{fig:Experimentalcure_cycles_combined}) to validate the model. The corresponding simulation data are subsequently used to assess the performance of the transfer learning approach. Because the final-time deformation is available experimentally, the R11 and R21 cases serve as validation benchmarks for the FiLM-DeepONet model to evaluate the accuracy of deformation-history predictions enabled by transfer learning.

\begin{figure}[!tbh]
    \centering
    \begin{subfigure}[t]{0.90\textwidth}
        \centering
        \includegraphics[width=\textwidth]{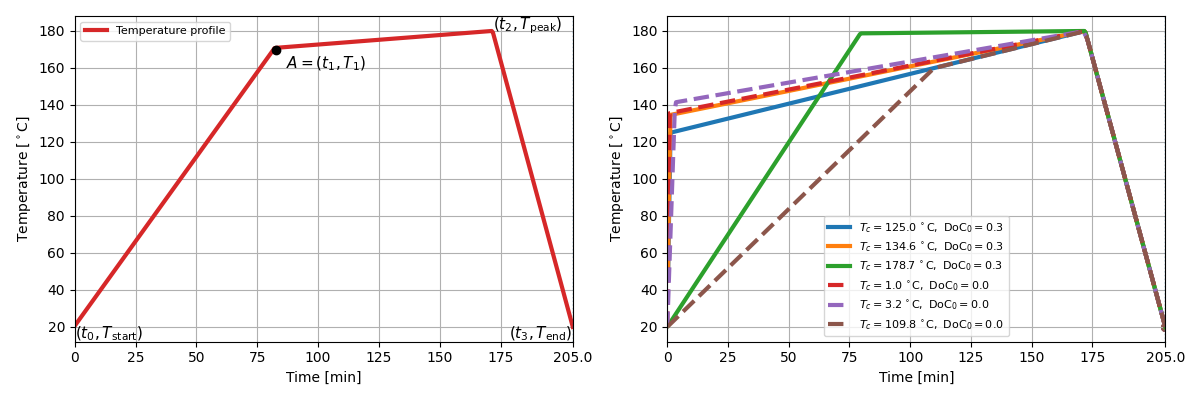}
    \caption{Parametrization of the cure temperature profile with an intermediate control point $A = (t_1, T_1)$ and fixed start $(t_0 = 0.333,\; T_{\text{start}} = 20.000)$, peak $(t_2 = 171.658,\; T_{\text{peak}} = 179.905)$, and end $(t_3 = 205.000,\; T_{\text{end}} = 20.000)$ temperatures, together with imposed temperature histories at selected temperature coordinates $T_c$ for cure cycles used to generate the simulation data in ABAQUS.}
        \label{fig:simulation_temp_opt}
    \end{subfigure}

    \vspace{0.8em}

    \begin{subfigure}[t]{0.99\textwidth}
        \centering
        \includegraphics[width=\textwidth]{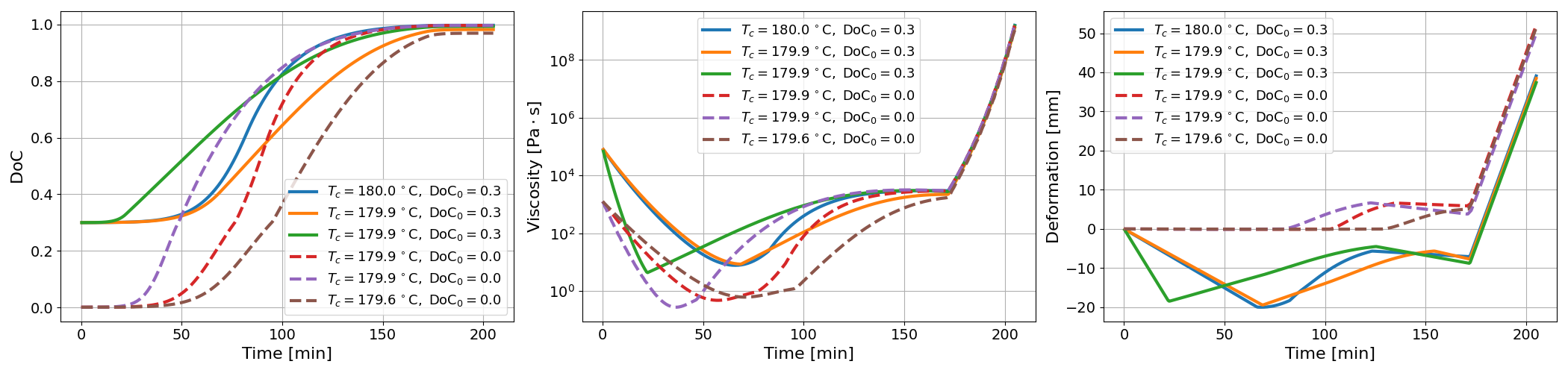}
        \caption{DoC and viscosity profiles from the thermo--chemical simulation, together with the corresponding deformation from the stress--deformation simulation, for $\mathrm{DoC}_0 = 0.3$ and $\mathrm{DoC}_0 = 0.001$.}
        \label{fig:simulation_outputs}
    \end{subfigure}

    \caption{Simulation data for a laminate subjected to a cure cycle. The top panel combines a parametrized cure temperature profile with an intermediate control point $A = (t_1, T_1)$ and fixed start, peak, and end temperatures, and  the corresponding imposed temperature histories at different temperature coordinates $T_c$ used in simulation data. The bottom panel shows the simulated evolution of DoC, viscosity, and deformation for two initial degrees of cure, $\mathrm{DoC}_0 = 0.3$ and $\mathrm{DoC}_0 = 0.001$.}
    \label{fig:simulation_data}
\end{figure}

As shown in Figure~\ref{fig:simulation_data}(a), a family of simulated temperature profiles is generated by varying an
intermediate temperature coordinate $A=(t_1, T_1)$ between the fixed start point
$(t_0, T_{\mathrm{start}}) = (0.333, 20)$ and peak point $(t_2, T_{\mathrm{peak}}) = (171.658, 179.905)$,
while keeping the end point $(t_3, T_{\mathrm{end}}) = (205, 20)$ fixed. This intermediate
coordinate controls both the duration and the temperature level of the dwell during the heating
stage. Figure~\ref{fig:simulation_data}(b) shows the corresponding histories of DoC, viscosity, and deformation
for different choices of the temperature coordinate $A$, illustrating the influence of $\mathrm{DoC}_{0}$ on the deformation evolution.

\section{Operator Learning}\label{sec:Operator-learning}
Operator learning provides a machine learning framework for approximating mappings between function spaces rather than finite-dimensional vectors~\cite{lu2019deeponet,lu2024bridging,shih2024transformers,kovachki2023neural}. The goal is to learn solution operators that map input functions, such as initial or boundary conditions, to output functions, for example the corresponding solutions of partial differential equations.
A prototypical architecture for this task is DeepONet~\cite{lu2019deeponet}, which employs a branch–trunk structure, the branch network encodes the input function, while the trunk network parameterizes the dependence on spatial or temporal query points.

\subsection{DeepONet}\label{sec:DeepONet}

DeepONet approximates nonlinear operators between function spaces using two neural networks in parallel, a \emph{branch} network and a \emph{trunk} network. In its general formulation, the branch network encodes the input function \( f \) by sampling it at a fixed set of sensor locations \( \{\bm{x}_1, \bm{x}_2, \ldots, \bm{x}_k\} \). The sampled values,
\[
\big(f(\bm{x}_1), f(\bm{x}_2), \ldots, f(\bm{x}_k)\big),
\]
are mapped through the branch network to produce a latent feature vector \( \mathbf{b} \in \mathbb{R}^p \). For a batch of \( m \) input functions, this results in a matrix \( B \in \mathbb{R}^{m \times p} \), with each row corresponding to one input sample.
The trunk network receives as input a spatial or temporal query point \( \bm{x} \in \mathbb{R}^d \) at which the operator output is to be evaluated and produces a corresponding latent representation \( \boldsymbol{\phi}(\bm{x}) \in \mathbb{R}^p \). For \( n \) evaluation points, the trunk network outputs a matrix \( \Phi \in \mathbb{R}^{n \times p} \).
The operator output \( \mathcal{G}(f_i) \) evaluated at a query point \( \bm{x}_j \) is obtained through the dot–product contraction of the branch and trunk outputs:
\begin{equation}
    \mathcal{G}(f_i)(\bm{x}_j)
    \approx
    \sum_{l = 1}^{p}
    B_{il}(f_i; \theta_B)\,\Phi_{jl}(\bm{x}_j; \theta_T),
    \label{eq:deeponet_output}
\end{equation}
where \( \theta_B \) and \( \theta_T \) denote the trainable parameters of the branch and trunk networks, respectively.

In this work, DeepONet is configured to learn the mapping from the imposed temperature profile \( \mathbf{T} \) of a cure cycle to the DoC, viscosity, and deformation history of the laminate. The branch network takes the temperature field \( \mathbf{T} \) as input, represented by samples of the cure-cycle temperature profile for different temperature coordinates $T_c$ (see Figure~\ref{fig:simulation_data}(a)),
\[
T_c = \{T(t_1), T(t_2), \ldots, T(t_k)\},
\]
which serve as a discretized representation of the input function. The trunk network takes the query time \( t \) as input and, for a set of evaluation times \( \{t_1, t_2, \ldots, t_n\} \), produces the corresponding latent basis functions \( \Phi(t_j) \), from which the time-dependent outputs are constructed.

\subsection{FiLM-enhanced DeepONet}\label{sec:FiLM-DeepONet}

As described in Section~\ref{sec:DeepONet}, the standard DeepONet architecture learns an operator by combining a branch network, which encodes the input function, with a trunk network, which represents the evaluation coordinates. However, in the present problem, the laminate response depends not only on the applied temperature history but also on $\mathrm{DoC}_{0}$. As illustrated in Figure~\ref{fig:simulation_data}(b), variations in $\mathrm{DoC}_{0}$ significantly influence the evolution of $\mathrm{DoC}$, viscosity, and deformation.

To incorporate this additional physical dependence, we augment the branch network with Feature-wise Linear Modulation (FiLM), resulting in a DeepONet architecture. FiLM provides a mechanism for conditioning intermediate branch features on an auxiliary variable—in this case, \( DoC_0 \)—allowing the operator representation to adapt dynamically to different initial material states.
For a hidden-layer activation \( h \) in the branch network, the FiLM transformation is given by
\[
h' = \gamma(DoC_0)\, h + \beta(DoC_0),
\]
where \( \gamma(DoC_0) \) and \( \beta(DoC_0) \) are learned affine functions of the conditioning input. More generally, for a vector of hidden features \( \mathbf{h} \) and conditioning input \( \mathbf{c} \),
\[
\text{FiLM}(\mathbf{h} \mid \mathbf{c})
    = \gamma(\mathbf{c}) \odot \mathbf{h} + \beta(\mathbf{c}),
\]
with \( \odot \) denoting element-wise multiplication. Through this modulation, the branch network produces a representation of the temperature history that is explicitly conditioned on $\mathrm{DoC}_{0}$.

In the present implementation, FiLM is applied to each hidden layer of the branch network, while the output layer is not modulated. The branch network consists of three hidden layers of width 20, followed by an output layer of dimension \(3G\). Thus, for each hidden layer \(\ell\), the FiLM modulation parameters \(\gamma^{(\ell)}(DoC_0)\) and \(\beta^{(\ell)}(DoC_0)\) are 20-dimensional feature-wise vectors whose dimensions match the hidden-layer width. These modulation vectors are generated from the scalar conditioning input \(DoC_0\) through separate affine transformations for each layer,
\[
\gamma^{(\ell)}(DoC_0)=DoC_0 W_{\gamma}^{(\ell)}+b_{\gamma}^{(\ell)}, \qquad
\beta^{(\ell)}(DoC_0)=DoC_0 W_{\beta}^{(\ell)}+b_{\beta}^{(\ell)}.
\]
The FiLM parameters are therefore layer-specific and are not shared across hidden layers.

In the Feature-wise Linear Modulation DeepONet used here, the branch network receives the sampled temperature profile \( \mathbf{T} \) together with \( DoC_0 \). 
Its output is partitioned into three latent vectors corresponding to the three predicted quantities:
\[
\mathbf{h}_{\text{branch}}(\mathbf{T}, d_0; \theta_{\text{branch}}) 
    = \big[\, \mathbf{h}_d,\; \mathbf{h}_v,\; \mathbf{h}_\varepsilon \,\big] 
    \in \mathbb{R}^{3G},
\]
where \( \mathbf{h}_d \), \( \mathbf{h}_v \), and \( \mathbf{h}_\varepsilon \in \mathbb{R}^G \) encode the operator representations for DoC, viscosity, and deformation, respectively.
The trunk network takes the query time \( t \) as input and produces a shared latent basis:
\[
\mathbf{h}_{\text{trunk}}(t; \theta_{\text{trunk}}) \in \mathbb{R}^G.
\]

Following the DeepONet formulation, the predicted outputs are obtained by inner products between the FiLM-modulated branch features and the trunk basis:
\[
\hat{d}(t) = \mathbf{h}_d^\top \mathbf{h}_{\text{trunk}}(t), \qquad
\hat{v}(t) = \mathbf{h}_v^\top \mathbf{h}_{\text{trunk}}(t), \qquad
\hat{\varepsilon}(t) = \mathbf{h}_\varepsilon^\top \mathbf{h}_{\text{trunk}}(t).
\]

The FiLM-conditioned branch remains shared across all three predicted quantities; that is, the modulation is applied to the shared branch representation before the final branch output is partitioned into separate latent vectors for DoC, viscosity, and deformation. Thus, FiLM is not implemented as three separate output-specific conditioning blocks, but rather as a unified conditioning mechanism acting on the common latent representation.


The purpose of FiLM in the present framework is not primarily to improve accuracy over a standard DeepONet for a fixed initial state, but to provide an explicit conditioning mechanism through which the scalar initial degree of cure can modify the shared branch representation and thereby influence the predicted response histories. This is particularly important here because the same cure-temperature history can lead to different DoC, viscosity, and deformation trajectories depending on the initial cure state. Here, the initial degree of cure, $DoC_0$, is a scalar conditioning variable that affects the output response histories. Unlike the temperature history, it is not itself a function to be mapped by the operator, but rather an auxiliary parameter used to condition the branch representation.

\section{DeepONet-Based Framework for Modeling Cure-Induced Material Response}\label{sec:DeepONet-Based_results}

\begin{figure}[ht]
\centering
\includegraphics[width=0.995\textwidth]{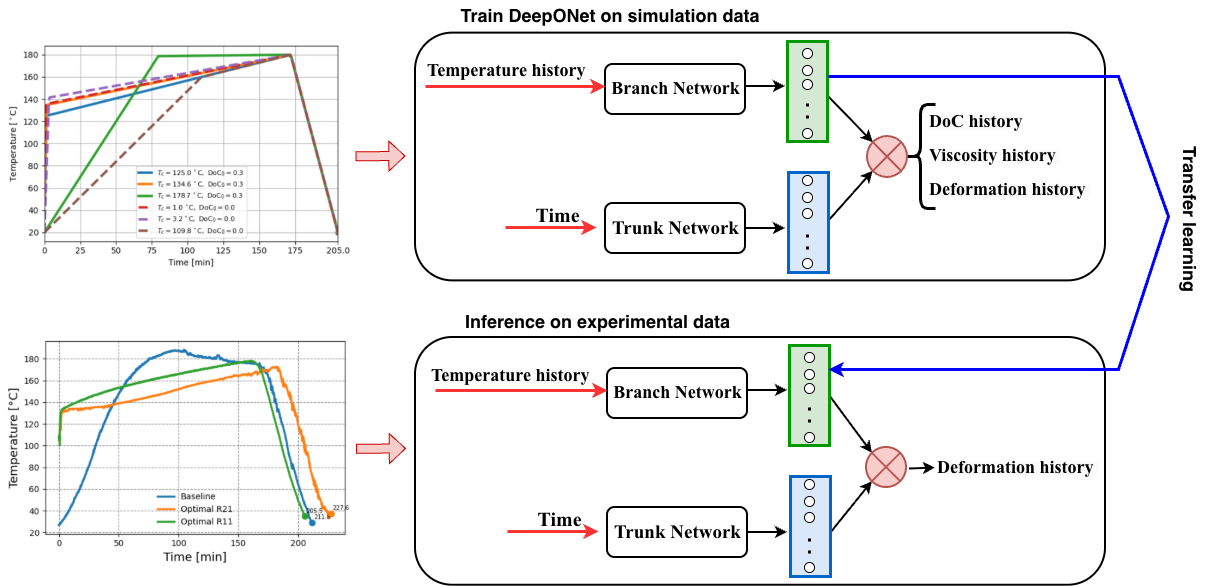}
\caption{Schematic of the DeepONet architecture for predicting DoC, viscosity, and deformation of a composite material during the cure process. In the first stage, DeepONet is trained on simulation data, the branch network takes the temperature history as input, while the trunk network takes time as input to predict the DoC, viscosity, and deformation histories. In the second stage, the trained DeepONet is used as a surrogate model, and transfer learning is applied by modifying the last layer of the branch network to satisfy the final deformation observed in the experimental data. An additional loss term is introduced to compare the predicted deformation with the experimental deformation, resulting in a final prediction of the deformation history under the experimental temperature profile.}
\label{fig:deeponet_Schematic}
\end{figure}

Figure~\ref{fig:deeponet_Schematic} presents a schematic of the DeepONet architecture used to predict the DoC, viscosity, and deformation of the composite during the cure process. In the first stage, DeepONet is trained on simulation data: the branch network takes the temperature profile as input, while the trunk network takes time as input, and their combination yields the full temporal histories of DoC, viscosity, and deformation.
The branch and trunk networks are shared across all three predicted quantities. Only at the final stage is the branch output partitioned into three latent vectors, corresponding to DoC, viscosity, and deformation. This shared architecture enables the model to learn a common representation of the cure history while preserving separate operator coefficients for each physical quantity.
In the second stage, the trained DeepONet is employed as a surrogate model for the experiments. The simulated temperature profiles are defined on a common time interval from \(0\) to \(205\) minutes, whereas the experimental temperature profiles span different time windows (e.g., up to approximately \(255\), \(211.6\), or \(227.6\) minutes, depending on the test) and are available only in limited quantity. Training a new network solely on this experimental dataset would be inefficient and prone to overfitting. Instead, a transfer-learning strategy is adopted: the weights learned from the simulation stage are retained, and only the last layer of the branch network is updated so that the predicted deformation matches the experimentally observed final deformation. An additional loss term penalizes the discrepancy between the predicted and measured deformation at the final time, yielding corrected deformation histories for each experimental temperature profile while still leveraging the rich operator learned from the simulations.

As illustrated in Figure~\ref{fig:deeponet_Schematic}, the branch network takes as input the discretized temperature field \( \mathbf{T} \), sampled at a set of temperature coordinates, together with $\mathrm{DoC}_{0}$, and outputs three sets of latent coefficients corresponding to the histories of DoC, viscosity, and deformation. The trunk network receives the time coordinates and produces a corresponding set of basis functions; the predicted histories are then obtained by combining the branch and trunk outputs. Both branch and trunk networks employ three hidden layers with 20 neurons each and hyperbolic tangent (\(\tanh\)) activation functions, followed by an output layer with 20 neurons. The model is trained on simulation data using the Adam optimizer with an exponentially decaying learning rate, initialized at \(10^{-3}\) and updated according to an exponential decay schedule, for up to \(10^{5}\) iterations with early stopping.
 The role of transfer learning becomes relevant only for experimental data, where a distribution shift exists between simulation and experiment. In this setting, the measured final deformation is used to adapt the simulation-trained model through a small parameter update, so that the prediction is corrected toward the experimental domain while preserving the physically meaningful trajectory learned from simulation. Thus, the full deformation history is not reconstructed from the final point alone; rather, it is obtained by refining a simulation-informed prior prediction.

\begin{figure}[ht]
    \centering
    \includegraphics[width=\textwidth]{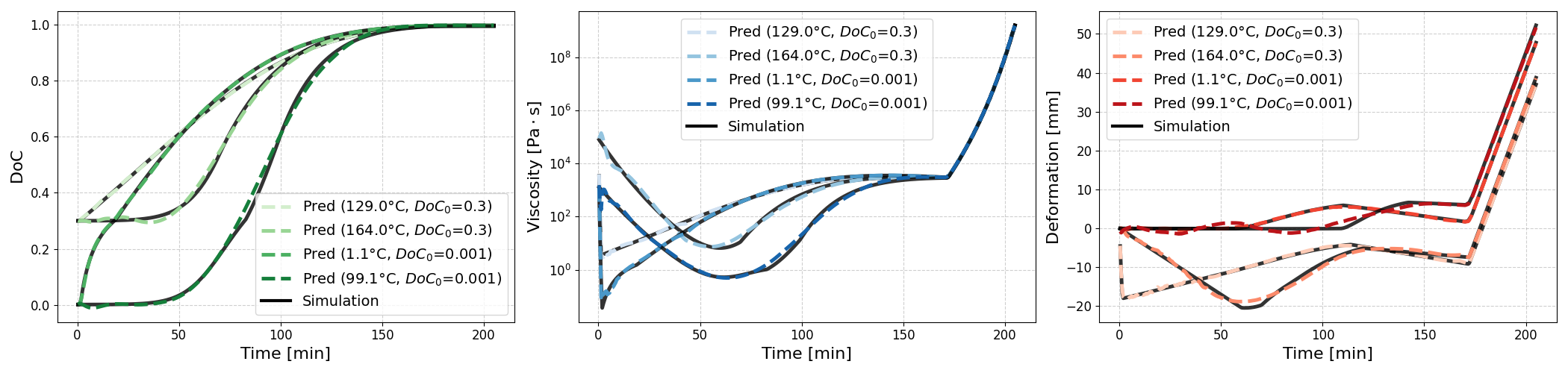}
    \caption{Comparison of FiLM-DeepONet predictions with true simulation results at three temperature coordinates, including DoC for two initial conditions ($\mathrm{DoC}_{0}=0.3$ and $\mathrm{DoC}_{0}=0.001$), viscosity, and deformation. $\mathrm{DoC}_{0}$ is provided as a conditioning input to the branch network.}
    \label{fig:predict_filmdeeponet}
\end{figure}

Figure~\ref{fig:predict_filmdeeponet} compares DeepONet predictions with reference simulation results for $\mathrm{DoC}$, viscosity, and deformation over time under four temperature profiles, using two initial conditions ($\mathrm{DoC}_{0}=0.001$ and $\mathrm{DoC}_{0}=0.3$). In the FiLM-enhanced architecture, $\mathrm{DoC}_{0}$ conditions the branch network, enabling the model to capture how the coupled thermochemical and mechanical response varies with the initial cure state. The close agreement between predictions and reference data demonstrates both the accuracy and generalization capability of the trained DeepONet. The rightmost panel further indicates that $\mathrm{DoC}_{0}$ strongly influences the final deformation pattern, justifying its inclusion as a conditioning variable.
As shown in the figure, viscosity grows approximately exponentially with time, leading to steep gradients and a wide dynamic range. Such highly skewed data can be challenging for neural networks to learn directly. To mitigate this, viscosity was log-transformed prior to training,
\[
\mu_{\text{log}} = \log(\mu + \varepsilon),
\]
where $\varepsilon = 10^{-8}$ is a small constant introduced for numerical stability. The figure also illustrates that FiLM enables the network to clearly distinguish responses corresponding to different initial degrees of cure.

\section{Uncertainty Quantification for DeepONet}\label{sec:Uncertainty_Quantification}

UQ is crucial for assessing the reliability of DeepONet predictions, particularly with noisy or limited data. By explicitly characterizing confidence in the model outputs, UQ enables more robust and trustworthy deployment of operator-learning frameworks in practical applications~\cite{iglesias2013ensemble}. In this section, we present UQ results for DeepONet, focusing on epistemic uncertainty and EKI. We demonstrate how transfer learning under uncertainty can predict deformation histories, together with associated uncertainty bounds, for the experimental temperature profiles.

\subsection{DeepONet Ensembles for Epistemic Uncertainty}
To quantify epistemic uncertainty, we train an ensemble of DeepONet models (specifically ten) using different random seeds for weight initialization, while keeping the training/validation split fixed. For each input, this ensemble yields multiple predictions, from which we compute the mean and standard deviation across seeds as the prediction and a measure of epistemic uncertainty, respectively \cite{psaros2023uncertainty}. In practice, we observe that the ensemble spread is relatively small, as independently initialized networks trained on the same data tend to converge to similar solutions.

\begin{figure}[ht]
    \centering
    \begin{subfigure}[b]{\textwidth}
        \centering
        \includegraphics[width=\textwidth]{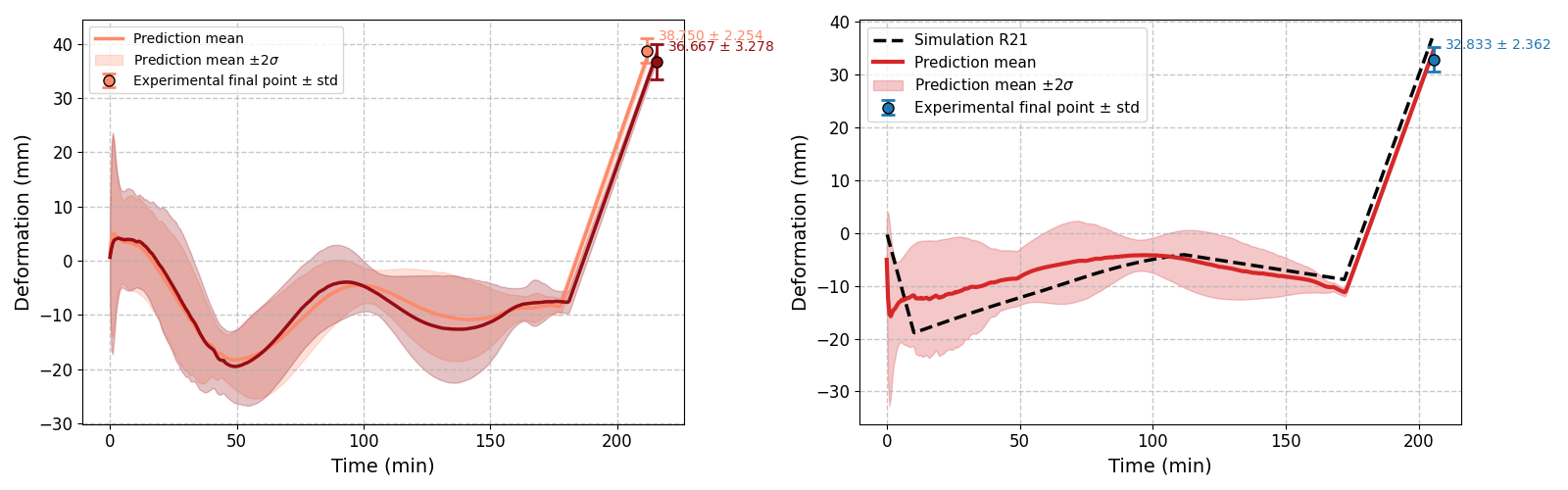}
        \label{fig:doc_pred}
    \end{subfigure}
\caption{Mean and standard deviation of DeepONet ensemble predictions for deformation.
(Left) Deformation histories for two baseline specimen experiments. The solid curves show the ensemble-averaged predictions (mean over ten independently trained models with different random seeds), and the shaded bands indicate the corresponding standard deviation. The experimental final deformation for each experiment is shown as a marker with an error bar, where the marker denotes the mean measured terminal deformation and the error bar indicates the corresponding experimental standard deviation. This illustrates the agreement between the transfer-learned predictions and the measured terminal values.
(Right) Deformation history for optimal R21, comparing the transfer-learned ensemble prediction against the corresponding simulation data. The shaded region again denotes the standard deviation across seeds, visualizing the epistemic uncertainty of the DeepONet ensemble. The final experimental deformation is also reported as a marker with an error bar, representing the mean measured value and its experimental standard deviation.}
\label{fig:Epistemic_Uncertainty}
\end{figure}

After this initial training, the ensemble is further refined via transfer learning by augmenting the loss function with a term that penalizes the mismatch between the predicted deformation at the final time step and the final deformation value reported in the experimental data. This encourages the network to remain consistent with the full deformation profile while accurately matching the terminal deformation.
Figure~\ref{fig:Epistemic_Uncertainty} (left) illustrates the deformation histories for two baseline specimen experiments. The curves show the ensemble-averaged deformation profile (mean over multiple seeds), with an uncertainty band corresponding to the ensemble standard deviation. The experimental final deformation is indicated as a single marker, demonstrating that the transfer-learned predictions closely match the reported terminal values.
The relatively large uncertainty near the beginning of the deformation histories arises mainly from the way uncertainty is estimated in the DeepONet ensemble and from the limited experimental supervision available during transfer learning. In Figure~\ref{fig:Epistemic_Uncertainty}, the uncertainty band is computed as the standard deviation across five independently initialized DeepONet models. Since the experimental data constrain the model only through the final deformation and do not provide the full time-resolved deformation history, the early-time portion of the trajectory is only weakly constrained. As a result, different ensemble members can produce noticeably different initial deformation paths while still matching the same terminal deformation, which leads to a larger spread near the beginning of the cure cycle.
Figure~\ref{fig:Epistemic_Uncertainty} (right) shows the predicted deformation history for the optimal R21 case obtained from the transfer-learned DeepONet ensemble, together with the corresponding simulation results used for validation. Since the full deformation history is not available from the experiments, the ensemble predictions are evaluated against the simulated deformation profile. The shaded region denotes the standard deviation across different seeds, providing a measure of the epistemic uncertainty associated with the DeepONet ensemble.

\subsection{Ensemble Kalman Inversion}\label{Ensemble-Kalman-Inversion (EKI)}
The ensemble Kalman filter (EnKF), originally introduced by Evensen in 1994 for data assimilation in time-dependent dynamical systems, has become a widely used tool owing to its robustness, ease of implementation, and strong numerical performance~\cite{evensen1994sequential}. Building on this foundation, iterative ensemble Kalman methods have been developed for solving inverse problems. In this context, the inferred solution is constrained to the subspace spanned by the initial ensemble, which allows for error estimates relative to the best approximation achievable within that subspace. Numerical studies have shown that this ensemble-based, derivative free strategy can attain accuracy comparable to classical least squares approaches, and in many cases close to the optimal approximation.

EKI is a gradient-free methodology specifically designed for inverse problems. Recent work~\cite{pensoneault2025uncertainty} applied EKI to the training of DeepONets and demonstrated that it achieves substantially faster convergence while providing more reliable uncertainty quantification than sampling based alternatives such as HMC. Motivated by these results, we integrate EKI into our operator learning framework to obtain efficient and informative uncertainty estimates for DeepONet.

\begin{figure}[!tbh]
    \centering
    \begin{subfigure}[t]{0.9\textwidth}
        \centering
        \includegraphics[width=\textwidth]{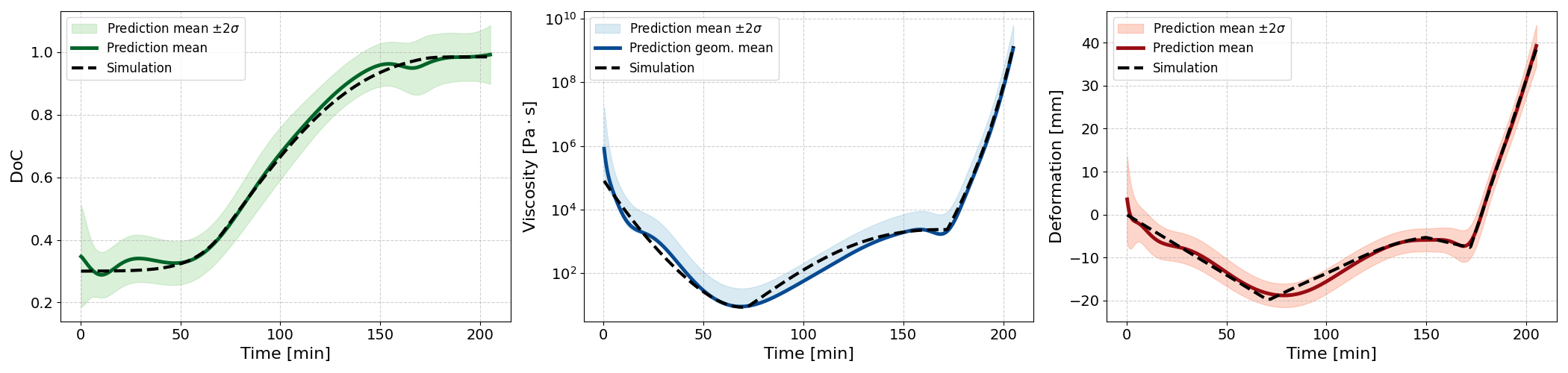}
        \caption{Comparison of ensemble predictions for DoC, viscosity, and deformation obtained with EKI against the corresponding simulation data.}
        \label{fig:EKI_uncertainty_a}
    \end{subfigure}

    \vspace{0.4cm}

    \begin{subfigure}[t]{0.9\textwidth}
        \centering
        \includegraphics[width=\textwidth]{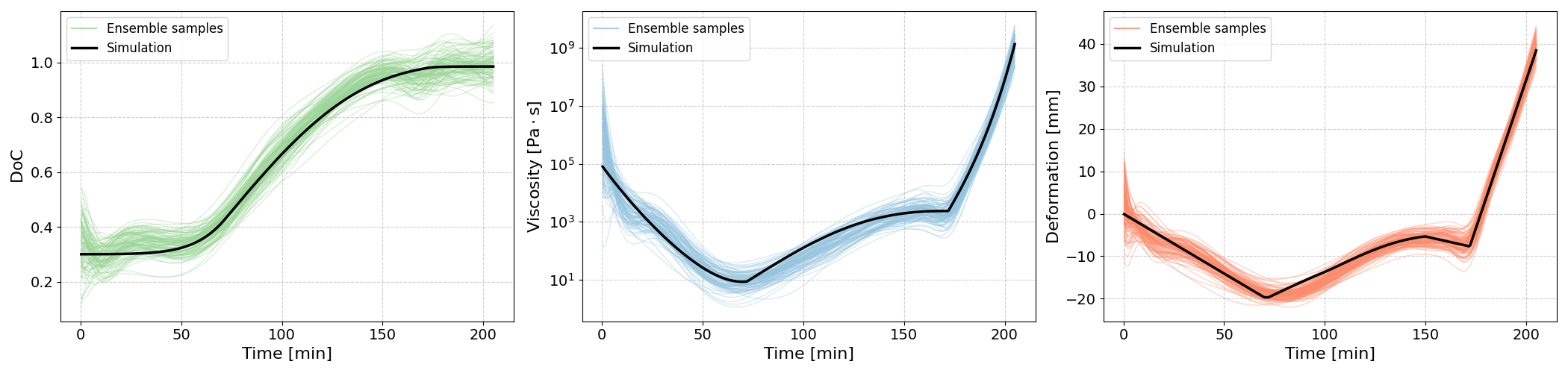}
        \caption{Ensemble sample trajectories for DoC, viscosity, and deformation, with solid lines indicating the reference simulation responses.}
        \label{fig:EKI_uncertainty_b}
    \end{subfigure}

    \caption{Prediction uncertainty and ensemble behavior for DoC, viscosity, and deformation using EKI. 
    (a) Uncertainty bands showing ensemble mean predictions (solid lines), reference simulation responses (dashed lines), 
    and shaded regions corresponding to one standard deviation of the ensemble. 
    (b) Ensemble trajectory plots illustrating the spread of individual samples around the reference simulation responses (solid lines).}
    \label{fig:EKI_uncertainty_combined}
\end{figure}

In detail, we assume that the noisy observations are generated by
\begin{equation*}
    G_i^{(l)}
    = \mathcal{G}^\star(u^{(l)})(y_i^{(l)}) + \varepsilon_i^{(l)}, 
    \qquad \varepsilon_i^{(l)} \sim \mathcal{N}(0, \sigma_l^2 I),
\end{equation*}
where $\mathcal{G}^\star$ denotes the (unknown) ground-truth operator and
$G_{i}^{(l)}$ is the noisy observation at the $i$-th query location for the
$l$-th input function. The noise variables $\varepsilon_i^{(l)}$ are
independent zero-mean Gaussian with noise level $\sigma_l$. For the $l$-th
input function $u^{(l)}$ we collect the corresponding noisy outputs in
$\mathcal{D}^{(l)} = \{ G^{(l)}_{i} \}_{i=1}^{P}$, where $G^{(l)}_{i}$ denotes
the noisy output at $P$ query locations. By stacking the datasets for all $N$
inputs we obtain the full noisy dataset
$\mathcal{D} = \{ \mathcal{D}^{(l)} \}_{l=1}^{N}$.
We assume that all observations are independent and identically distributed
(i.i.d.) and follow a Gaussian likelihood. To approximate the true operator we
introduce a DeepONet operator
$\mathcal{G}_{\boldsymbol{\theta}} \approx \mathcal{G}^\star$ parameterized
by $\boldsymbol{\theta}\in \mathbb{R}^{N_{\theta}}$. The operator learning
task then becomes a Bayesian inverse problem with likelihood
\begin{equation}
p(\mathcal{D} \mid \boldsymbol{\theta})
= \prod_{l=1}^{N} \prod_{i=1}^{P}
\frac{1}{\sqrt{2\pi\sigma_{l}^{2}}}
\exp\left(
  -\,\frac{\bigl(G^{(l)}_{i}
     - \mathcal{G}_{\boldsymbol{\theta}}(u^{(l)})(y^{(l)}_{i})\bigr)^{2}}
          {2\sigma_{l}^{2}}
\right).
\label{eq:likelihood}
\end{equation}
Assuming a standard multivariate Gaussian prior on $\boldsymbol{\theta}$ and
applying Bayes' formula, the posterior distribution is given by

\begin{equation*}
    p(\boldsymbol{\theta}\mid\mathcal{D})
    \propto p(\mathcal{D}\mid\boldsymbol{\theta})\,p(\boldsymbol{\theta}).
\end{equation*}

To approximate this posterior distribution, EKI considers evolving an ensemble of particles according to the dynamical system

\begin{equation*}
    \begin{split}
        \boldsymbol{\theta}_t &= \boldsymbol{\theta}_{t-1} + \boldsymbol{\xi}_t,
        \qquad \boldsymbol{\xi}_t \sim \mathcal{N}(0, Q),\\[1mm]
        \mathbf{y}_t &= \mathcal{F}(\boldsymbol{\theta}_t) + \boldsymbol{\zeta}_t,
        \qquad \boldsymbol{\zeta}_t \sim \mathcal{N}(0, R),
    \end{split}
\end{equation*}

where $\boldsymbol{\xi}_t$ is an artificial process noise with covariance
matrix $Q\in \mathbb{R}^{N_{\theta}\times N_{\theta}}$, and
$\mathcal{F}(\boldsymbol{\theta}_t)$ denotes the forward map
obtained by evaluating the DeepONet operator on the fixed training inputs,
\[
\mathcal{F}(\boldsymbol{\theta}_t)
= \mathrm{vec}\left(
  \left\{
    \mathcal{G}_{\boldsymbol{\theta}_t}(u^{(l)})(y_i^{(l)})
  \right\}_{l=1,\dots,N;\;i=1,\dots,P}
\right)
\in \mathbb{R}^{NP},
\]
while $\boldsymbol{\zeta}_t$ represents the (flattened) observation noise with
covariance $R\in \mathbb{R}^{NP\times NP}$. Given an ensemble of initial
particles $\{\boldsymbol{\theta}_{0}^{(j)}\}_{j=1}^{J}$ and using the
marginals of the joint Gaussian distribution, the EKI update for the ensemble
$\{\boldsymbol{\theta}_t^{(j)}\}_{j=1}^J$ reads
\begin{align}
    \hat{\boldsymbol{\theta}}_t^{(j)}
    &= \boldsymbol{\theta}_{t-1}^{(j)} + \boldsymbol{\xi}_t^{(j)},
    & \boldsymbol{\xi}_t^{(j)} &\sim \mathcal{N}(0, Q), \tag{2.10}\\
    \hat{\mathbf{y}}_t^{(j)}
    &= \mathcal{F}\bigl(\hat{\boldsymbol{\theta}}_t^{(j)}\bigr), \tag{2.11}\\
    \boldsymbol{\theta}_t^{(j)}
    &= \hat{\boldsymbol{\theta}}_t^{(j)}
       + C_t^{\hat{\theta}y}\bigl(C_t^{yy} + R\bigr)^{-1}
         \bigl(\mathbf{y} - \hat{\mathbf{y}}_t^{(j)} + \boldsymbol{\zeta}_t^{(j)}\bigr),
    & \boldsymbol{\zeta}_t^{(j)} &\sim \mathcal{N}(0, R), \tag{2.12}
\end{align}
where $C_t^{yy}$ and $C_t^{\hat{\theta}y}$ are the empirical covariance
matrices, defined as
\begin{align}
    C_t^{yy} &= Y_t Y_t^{T}, 
     C_t^{\hat{\theta}y} = \Theta_t Y_t^{T}, \tag{2.13}\\
    Y_t &= \frac{1}{\sqrt{J-1}}
    \bigl[
        \hat{\mathbf{y}}_t^{(1)} - \bar{\mathbf{y}}_t,\,
        \ldots,\,
        \hat{\mathbf{y}}_t^{(J)} - \bar{\mathbf{y}}_t
    \bigr], \tag{2.14}\\
    \Theta_t &= \frac{1}{\sqrt{J-1}}
    \bigl[
        \hat{\boldsymbol{\theta}}_t^{(1)} - \bar{\boldsymbol{\theta}}_t,\,
        \ldots,\,
        \hat{\boldsymbol{\theta}}_t^{(J)} - \bar{\boldsymbol{\theta}}_t
    \bigr]. \tag{2.15}
\end{align}
Here, $\bar{\boldsymbol{\theta}}_t$ and $\bar{\mathbf{y}}_t$ denote the sample
means of the prior ensembles $\{\hat{\boldsymbol{\theta}}_t^{(j)}\}_{j=1}^J$
and $\{\hat{\mathbf{y}}_t^{(j)}\}_{j=1}^J$ at iteration $t$, respectively.

\begin{figure}[!tbh]
    \centering
    {\small
    \makebox[0.45\textwidth][c]{\textbf{Predictions without transfer learning}}%
    \hfill
    \makebox[0.45\textwidth][c]{\textbf{Predictions with transfer learning}}\\[0.5ex]}
    \includegraphics[width=0.49\textwidth]{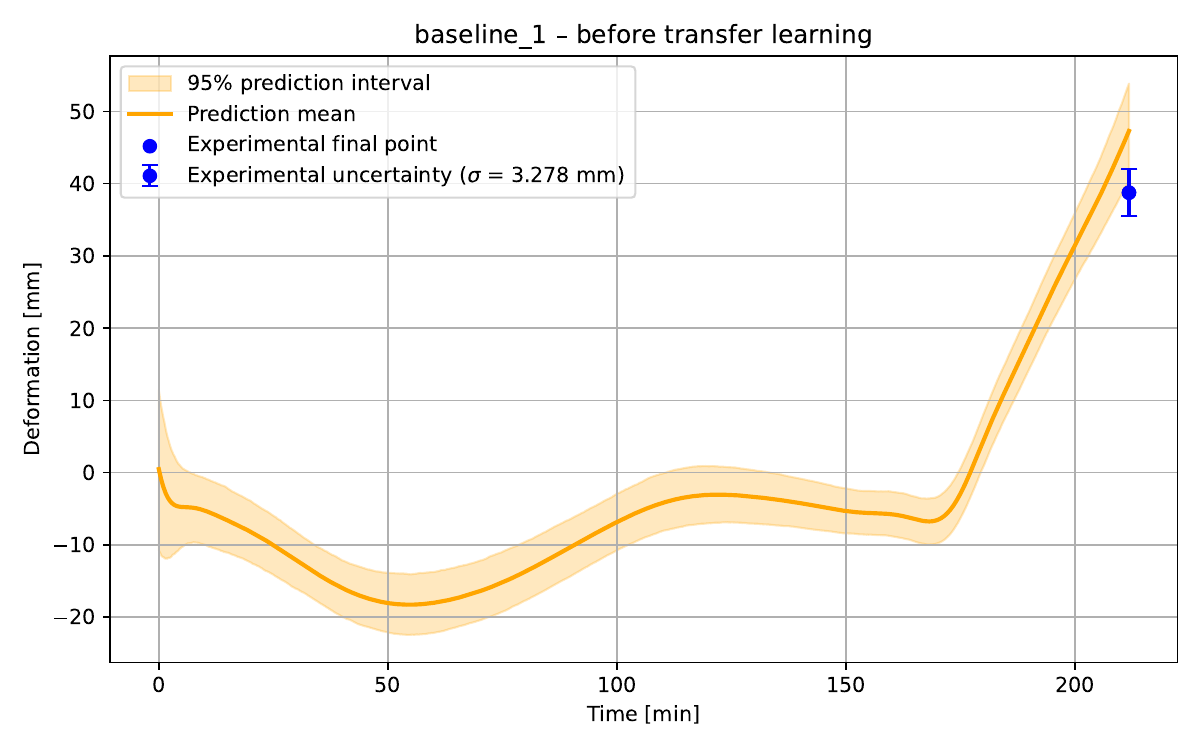}%
    \hfill
    \includegraphics[width=0.49\textwidth]{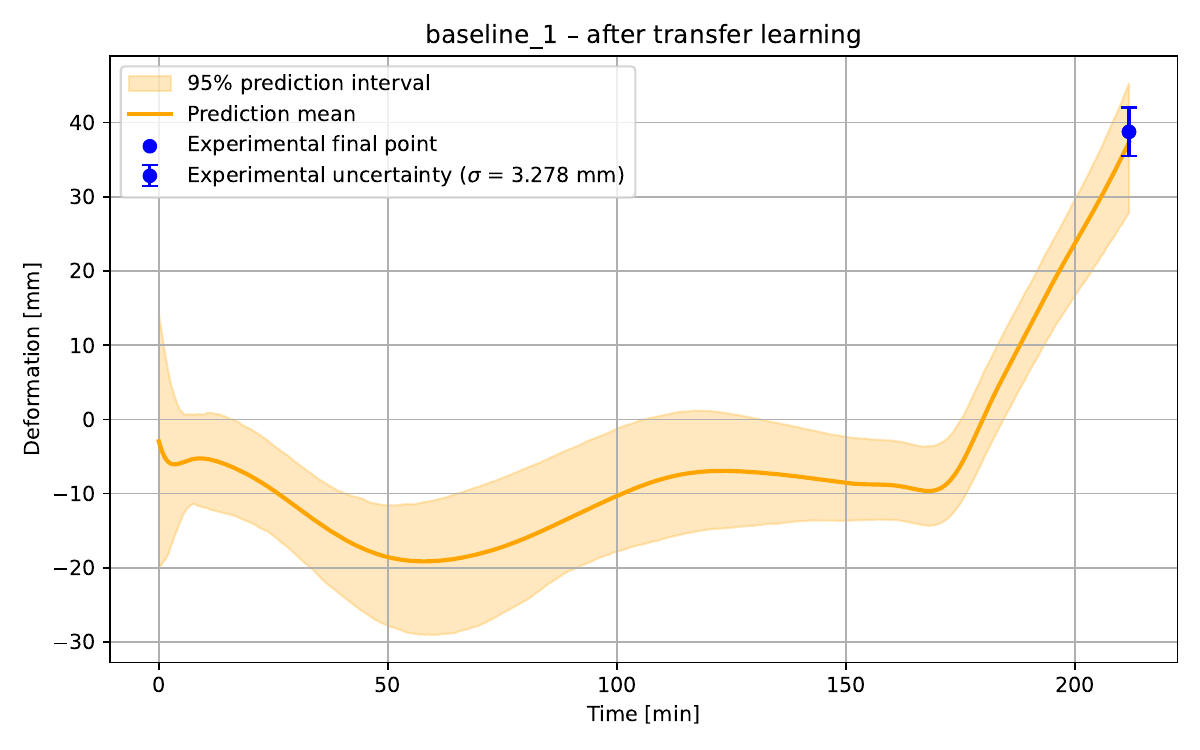}\\[0.5ex]
    \includegraphics[width=0.49\textwidth]{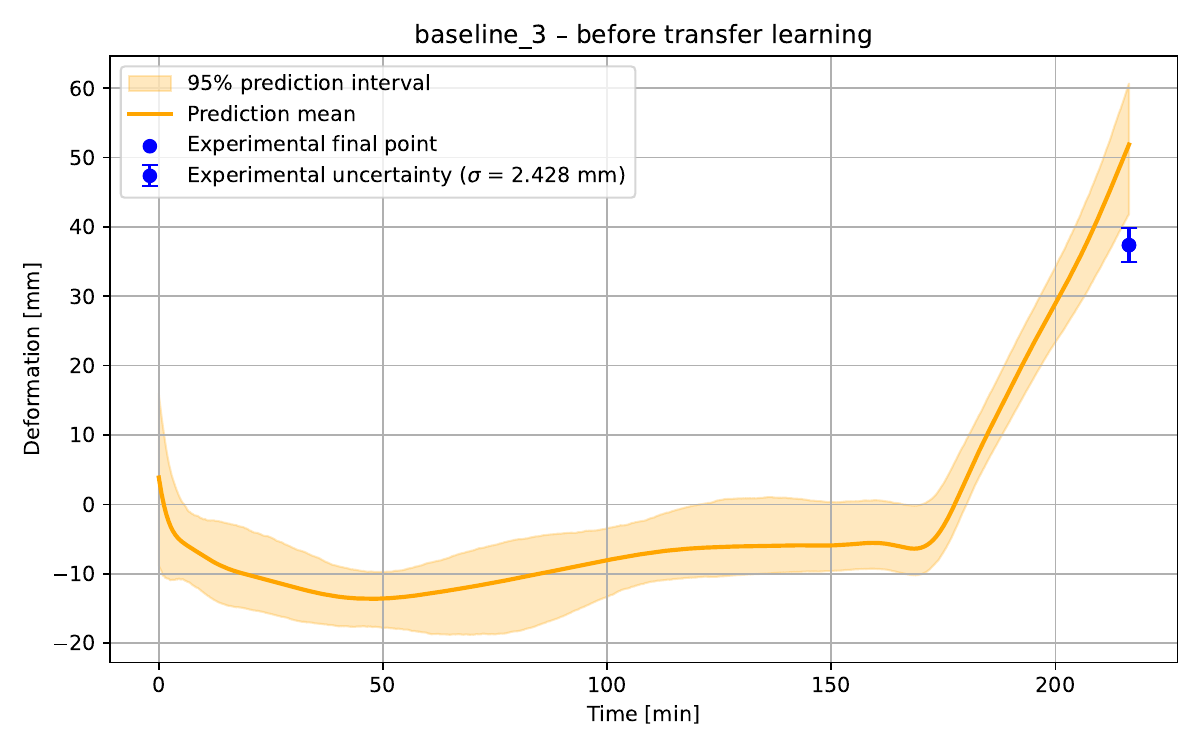}%
    \hfill
    \includegraphics[width=0.49\textwidth]{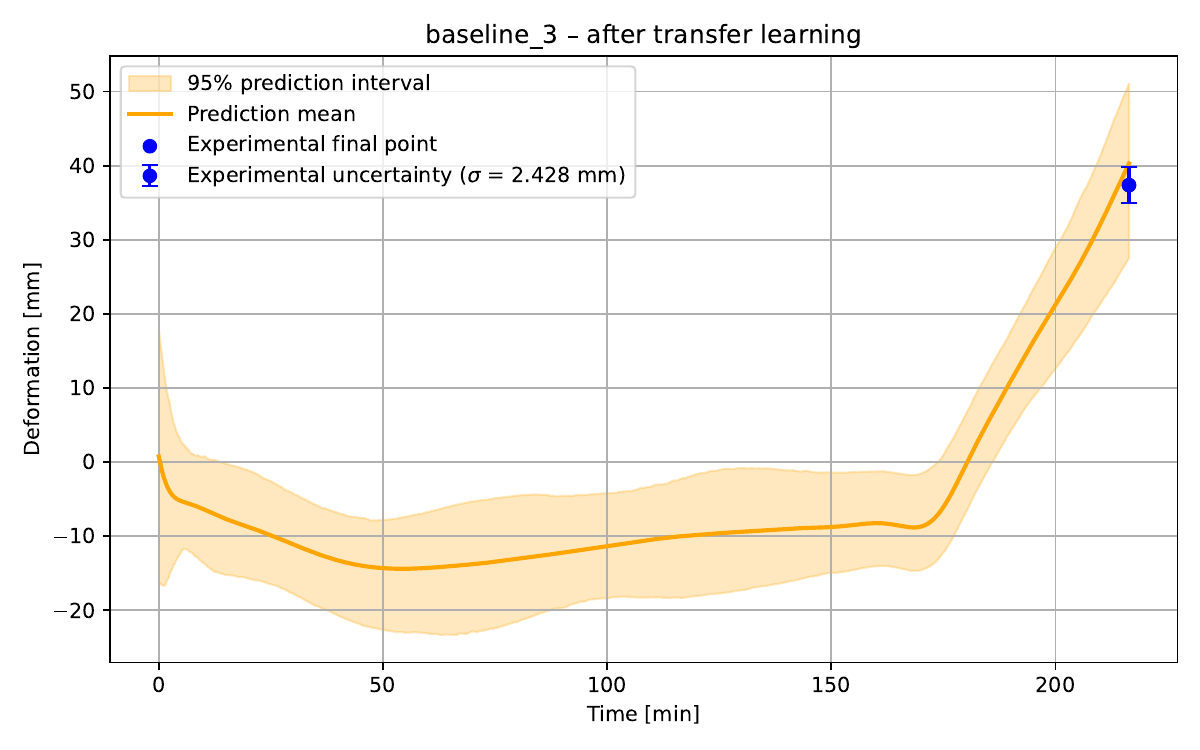}    
    \caption{Ensemble-based deformation predictions generated using EKI based on simulation data. 
Solid lines denote the ensemble mean predictions, and shaded regions represent the empirical 95\% prediction intervals obtained from the ensemble. 
Results are shown for two experimental baseline specimens: the left panels present predictions before transfer learning, while the right panels show predictions after transfer learning. 
Blue markers indicate the experimentally measured final deformation, and the corresponding error bars denote the associated experimental uncertainty.}
    \label{fig:EKI_ensemble_transfer_learning}
\end{figure}

In detail, we use an ensemble of size $J = 2000$ and run EKI for 1000 iterations. The standard deviation of the observation is chosen based on Table.\ref{tab:pid_validation}. To mitigate ensemble collapse, 
we also introduce artificial process noise with covariance $Q = 0.002 I$. 
The trunk and branch networks each contain two hidden layers with 10 neurons per layer, 
and their output layers produce 30-dimensional feature vectors. 
This compact network design substantially reduces the total number of trainable parameters.

Figure~\ref{fig:EKI_uncertainty_combined} summarizes the resulting predictions of DoC, viscosity, and deformation on the simulation test dataset together 
with their associated uncertainty estimates. Panel~(a) displays uncertainty bands, 
where solid lines denote the ensemble mean, dashed lines indicate the reference 
simulation responses, and shaded regions correspond to one standard deviation 
of the ensemble. Panel~(b) shows ensemble ``spaghetti'' trajectories, illustrating 
the spread of individual samples around the reference responses (solid lines). 
We observe that the prediction accuracy is high and that the confidence intervals 
consistently cover the ground truth trajectories, indicating that the proposed 
EKI–DeepONet framework yields meaningful and well-calibrated uncertainty quantification.

\begin{figure}[!tbh]
    \centering
    \includegraphics[width=0.85\textwidth]{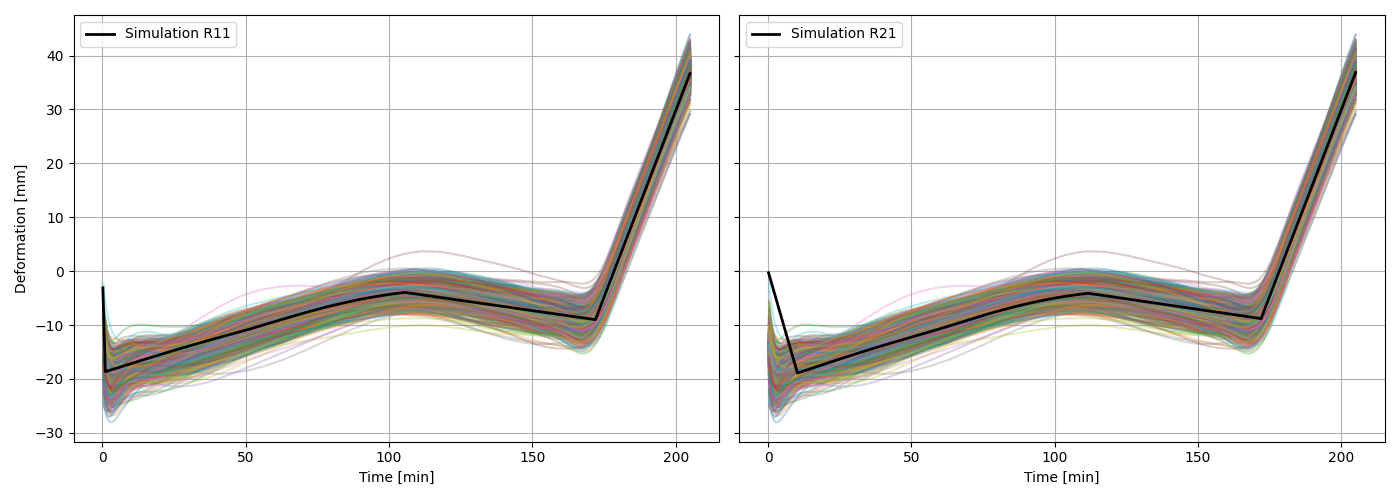}
\caption{Ensemble samples generated using EKI for the deformation in experimental optimal cases R11 and R21, shown alongside the corresponding simulation results for comparison. The solid black line represents the simulation data used for validation, while the colored lines denote the ensemble samples generated by EKI.}
    \label{fig:EKI_ensemble_100samples}
\end{figure}

For inference on the experimental data, we employ a transfer learning strategy. 
It is important to emphasize that the deformation history is not reconstructed 
solely from the final measurement. Instead, the prediction is primarily governed 
by the strong prior learned from the simulation-trained DeepONet, which captures 
the underlying deformation dynamics. Due to the distribution shift between simulation and experimental data, the 
predicted final deformation may deviate from the observed value. To address this, 
we use the final experimental measurement to perform a mild correction of the 
prediction through transfer learning, rather than learning the full trajectory 
from scratch. Since the experimental loss function is defined only at the final time point, 
the risk of overfitting is particularly high. To mitigate this, we fine-tune 
only the parameters in the last layer of the branch network, while keeping all 
other parameters fixed at the values obtained from training on the simulation data. 
Moreover, we employ Tikhonov-regularized EKI with a regularization parameter of 
$0.1$ to further stabilize the update and control overfitting.
Figures~\ref{fig:EKI_ensemble_transfer_learning} and~\ref{fig:EKI_ensemble_100samples} 
show the predicted deformation on the experimental dataset, before and after 
transfer learning, for the two case studies (two baseline specimens). 
After transfer learning, the prediction at the final time is significantly 
closer to the experimental ground truth, while the uncertainty along the earlier 
portions of the trajectory increases, resulting in more conservative and 
better-posed predictive uncertainty.

Figure~\ref{fig:EKI_ensemble_100samples} shows ensemble samples generated using EKI for the deformation in experimental cases R11 and R21, plotted together with the corresponding simulation results for comparison. This figure illustrates the ability of the network to generate predictive sample trajectories. To assess performance, we validate the ensemble predictions against the simulation data for cases R11 and R21. The close agreement between the simulation curves and the EKI-generated ensembles indicates that the method provides a reasonable match for these cases.

\section{Optimization of the cure temperature profile}\label{sec:Optimization}

As illustrated in Figure~\ref{fig:simulation_data}, the cure process is parameterized by the time and temperature of an intermediate point \( A = (t_1, T_1) \), which determines how long and at what temperature the resin is heated during the dwell stage.
In the optimization problem, the start, peak, and end temperatures and their corresponding times are fixed. The sole design variable is the intermediate point \( A = (t_1, T_1) \). By appropriately choosing \( (t_1, T_1) \), the intermediate heating segment of the cure cycle is reshaped to reduce the final process-induced deformation while still achieving full cure.

Let \( DoC(t;T) \) denote the DoC and \( u(t;T) \) the magnitude of process-induced deformation predicted by the trained DeepONet surrogate under a given temperature schedule \( T(\cdot) \). The optimization problem is

\begin{equation}
\label{eq:optimization_problem}
\begin{aligned}
\min_{t_1,T_1}\quad & J(t_1,T_1) := \big\|u\!\left(t_3;\,T(\cdot;t_1,T_1)\right)\big\|,\\[4pt]
\text{s.t.}\quad
& \text{DoC} \!\left(t_3;\,T(\cdot;t_1,T_1)\right) \;\ge\; 0.990 \quad \text{(full cure)},\\
& t_0+\Delta t \;\le\; t_1 \;\le\; t_2-\Delta t,\\
& T_{\text{start}} \;\le\; T_1 \;\le\; T_{\text{peak}},\\
& m_1(t_1,T_1) \;>\; m_2(t_1,T_1) \;>\; 0,
\end{aligned}
\end{equation}

where \( A = (t_1,T_1) \) are the design variables that control the shape of the intermediate segment of the cure schedule. In practice, we explore a wide range of candidate points \(A\) within the admissible bounds, construct the corresponding temperature profiles \( T(\cdot;t_1,T_1) \), and pass each profile through DeepONet to evaluate \( u(t_3;T(\cdot;t_1,T_1)) \) and \( DoC(t_3;T(\cdot;t_1,T_1)) \). The constrained optimization problem is then solved over \(A = (t_1,T_1) \) to identify a cure schedule that minimizes the final deformation while satisfying the degree-of-cure requirement. 
The present surrogate-based optimization framework can also be viewed in the context of other recent machine-learning-assisted process-optimization studies. Schoenholz and Zobeiry~\cite{schoenholz2024accelerated} proposed a theory-guided probabilistic machine-learning approach based on limited experiments and Gaussian process regression to optimize layup and cure-cycle parameters for reducing process-induced deformation, thereby avoiding the need for full material characterization and repeated process simulations. Lavaggi et al.~\cite{2022_Lavaggi} developed theory-guided machine-learning surrogates for autoclave co-curing of sandwich composite structures and showed that the learned models could replace repeated physics-based simulations in the optimization loop with substantial computational savings. In contrast to these scalar-response optimization frameworks, the present work employs a DeepONet surrogate to learn the operator mapping from cure-temperature history to the full temporal histories of degree of cure, viscosity, and deformation, and then uses this surrogate for constrained optimization of modified non-isothermal cure profiles. Together with the close agreement between the present optimized cure profile and the trends reported in~\cite{18_Limaye2024}, these comparisons further support the efficiency and consistency of the proposed DeepONet-based optimization framework. 

\begin{figure}[!tbh]
    \centering
    \includegraphics[width=0.89\textwidth]{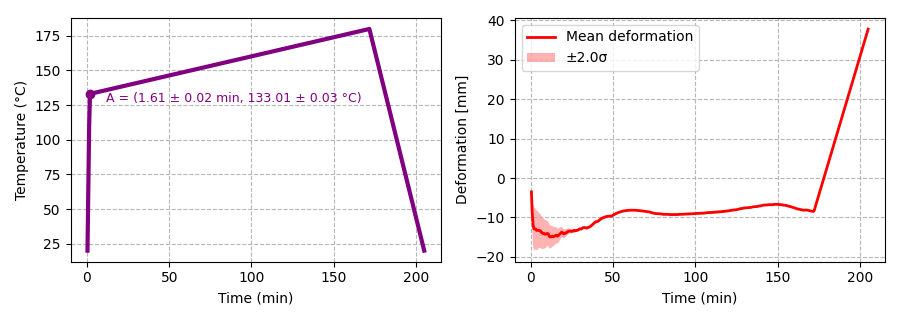}
    \caption{Result of the cure schedule optimization. The left panel shows the optimized temperature profile with the selected intermediate point \(A = (t_1, T_1)\), and the right panel shows the corresponding deformation history predicted by DeepONet for this temperature profile.}
    \label{fig:optimized_Temp_Def}
\end{figure}

Figure~\ref{fig:optimized_Temp_Def} illustrates the performance of DeepONet for optimizing the deformation under the constrained optimization problem~\eqref{eq:optimization_problem}. The predicted optimal cure schedule corresponds to the intermediate point
\(A = (t_1, T_1) \approx (1.61 \pm 0.02~\text{min},\, 133.01 \pm 0.03~^\circ\text{C})\).
It is worth noting that this DeepONet optimized temperature profile is very close to the optimal  profile previously obtained in~\cite{18_Limaye2024}  and aligns well with the R21 cure cycle implemented in Section~\ref{sec:Experimental_Materials_and_Methods}. Specifically, the optimized temperature profile obtained by DeepONet is very similar to the R21 cure profile obtained through simulation optimization, with the corresponding DeepONet final PID prediction of 37.81~mm. This is in close agreement with the simulation-based R21 PID value of 36.84~mm reported in Table~\ref{tab:pid_validation}. While a small difference remains relative to the experimentally measured final PID for the R21 cure cycle, the DeepONet-optimized result is nevertheless consistent with the simulation trend and shows a similar level of deviation from experiment as the simulation-based final PID prediction. Because the R21 cure cycle has already been experimentally implemented and assessed, this agreement provides indirect experimental validation for the optimized profile predicted by DeepONet. More broadly, these results indicate that further improving the fidelity of the physics-based simulation data used for DeepONet training is crucial for narrowing the remaining gap between AI model predictions and experimental measurements.

A useful comparison is~\cite{18_Limaye2024}, where non-isothermal cure-cycle parameters were optimized using a process model together with NSGA-II. In that work, the optimized cure cycles were obtained by varying the transition point between two ramp segments under a full-cure constraint, and the resulting deformation reduction was about 8--10\% relative to the baseline cure cycle. The modified cure profiles considered there are consistent with the family of cure cycles used to generate the training samples in the present study. Therefore, the close agreement between the optimized profile obtained here and the trends reported in~\cite{18_Limaye2024}, together with the final DeepONet-predicted deformation of 37.81~mm compared with the simulation-based R21 value of 36.84~mm, suggests that the DeepONet-based surrogate optimization is able to recover nearly the same optimum while avoiding repeated process-model evaluations during the search.


\section{Summary} \label{sec:Summary}

We have presented a data-driven framework for predicting and mitigating process-induced deformation (PID) in composite laminates by combining high-fidelity simulations, targeted experiments, and deep operator learning. PID arises from mismatches between fiber and matrix responses during curing, driven by thermal expansion and cure shrinkage, and is modeled here using a two-mechanism thermo-chemical and stress–deformation framework calibrated and validated against manufacturing trials. Comparisons between measured and simulated PID show that the model captures the dominant physics, with baseline predictions lying within the experimental scatter and optimized cure cycles achieving an 8–10\% reduction in deformation.

The validated model is then used to generate a diverse dataset of PID responses over a broad family of non-isothermal cure cycles. Using this dataset, we train a FiLM-enhanced DeepONet to predict the time histories of DoC, viscosity, and deformation for arbitrary temperature profiles, with Feature-wise Linear Modulation allowing the branch network to account for external parameters such as the initial DoC. To incorporate sparse experimental information, we apply transfer learning in which only the final layer of the pretrained network is updated using measured terminal deformation, while the underlying operator learned from simulation data is kept fixed.

Finally, an ensemble of DeepONet models is combined with Ensemble Kalman Inversion (EKI) to both quantify epistemic uncertainty and optimize the intermediate dwell point in the cure cycle, yielding temperature histories that minimize final deformation subject to a full-cure constraint. Overall, this framework illustrates how physics-based simulation, operator learning, and EKI-based uncertainty quantification can be integrated to construct reliable and data-efficient surrogates for composite cure process optimization, and it is readily extensible to other material systems as a foundation for robust cure-cycle design under uncertainty.  While the experimental standard deviations reported in Table~2 and shown in Figs.~9 and~11 provide a useful reference for the observed manufacturing scatter, the present DeepONet/EKI uncertainty should be understood mainly as model-based predictive uncertainty, not as a full decomposition of experimental, material, process, and model-form uncertainties.

As discussed in Section~\ref{sec:Experimental_Materials_and_Methods}, consolidation pressure is a key process parameter that would influence process induced deformation. A more exhaustive investigation focusing on consolidation pressure as a varying process parameter influencing PID in unbalanced unidirectional continuous fiber reinforced composites can be pivotal in advancing this work.

Furthermore, geometrical factors and layup configuration influence PID for composite components. A change in the geometry of the coupons or panels being manufactured from rectangular to square can change the profile of the cured composite laminate from a cylindrical to a saddle shape \cite{Ren2003}. This can influence the methodology for PID measurement and should be considered for such future investigations, where the geometry of the specimens is different from the one considered for this study. State-of-the-art technologies such as 3D scanning can be utilized to experimentally measure process induced deformation for saddle shaped profiles or more complex geometries \cite{kawagoe2022multiscale} such as L-brackets or thick composite beams with various cross-sections. This can form the basis for more detailed investigations to expand the generalizability of the framework developed as part of this study.

The methodology to predict PID based on process parameters, developed as part of this study, is applicable for unbalanced laminates and can be used as a tool to guide processing parameters’ identification (time-temperature cycle for curing) without relying on exhaustive constituent material characterization and physics-based simulation model development. Furthermore, it can reduce the physical trials needed to identify an optimal cure cycle since the model can provide much more suitable process parameters as a starting point for experiments. A limitation of the present study is that, although the final deformation is compared with experiments, the intermediate time histories of degree of cure, viscosity, and deformation remain simulation-informed quantities and are not yet directly validated through time-resolved experimental measurements.  

\section*{Acknowledgments}
This work was supported as part of the AIM for Composites, an Energy Frontier Research Center funded by the U.S. Department of Energy (DOE), Office of Science, Basic Energy Sciences (BES), under award \#DE-SC0023389. This research used computing resources provided by the Center for Computation and Visualization (CCV) at Brown University. 

\biboptions{sort&compress}
\bibliographystyle{elsarticle-num}
\bibliography{reference_clean}

\newpage
\end{document}